\documentclass[11pt,a4paper,onecolumn]{IEEEtran}
\usepackage[utf8]{inputenc}  
\usepackage{url}
\usepackage{amsmath}
\usepackage[makeroom]{cancel}
\usepackage{amsfonts}
\usepackage{amssymb}
\usepackage[pdftex]{graphicx}

\graphicspath{{../pdf/}{../jpeg/}}
\DeclareGraphicsExtensions{.pdf,.jpeg,.png}
\usepackage{pgf,tikz}
\usepackage{hyperref}
\usepackage{caption}
\usepackage{placeins}
\usepackage[left=2.25cm,right=2.25cm,top=2cm,bottom=2cm]{geometry}
\author{Tianshu Gao$^\dagger$, Mengbin Ye$^{\dagger,\star}$, and Robert Ackland$^\ddagger$
\thanks{$^\dagger$Centre for Optimisation and Decision Science, Curtin University, Perth, Australia}
\thanks{$^\ddagger$School of Sociology, Australian National University, Canberra, Australia}
\thanks{$^\star$Corresponding author: e-mail: mengbin.ye@curtin.edu.au  }
} 
\title{\huge {Multi-layer network analysis of deliberation in an online discussion platform: the case of Reddit}}
\usepackage{tabularx}
\usepackage{float}
\usepackage{amsthm}  
\usepackage{enumitem} 
\usepackage{bbm} 
\usepackage{cleveref}
\usepackage{bm}
\usepackage{multirow}
\usepackage{csvsimple}
\usepackage{colortbl}
\usepackage{booktabs}
\usepackage{csvsimple}
\usepackage{nomencl}

\usepackage{natbib}

\makenomenclature
\usepackage{subfigure}
\usepackage{array}

\renewcommand{\eqref}[1]{Eq.~(\ref{#1})}  

\newcommand{\vertiii}[1]{{\left\vert\kern-0.25ex\left\vert\kern-0.25ex\left\vert #1 
    \right\vert\kern-0.25ex\right\vert\kern-0.25ex\right\vert}} 

\date{}
\usepackage{hyperref} 
\hypersetup{
	citebordercolor = 1 1 1,
	linkbordercolor = 1 1 1,
	menubordercolor = 1 1 1,}

\begin{document}



\maketitle 

\begin{abstract}
This paper uses a multi-layer network model to study deliberation in online discussion platforms, focusing on the Reddit platform. The model comprises two layers: a discussion layer, which represents the comment-to-comment replies as a hierarchical tree, and an actor layer, which represent the actor-to-actor reply interactions. The interlayer links represent user-comment ownership. We further propose several different network metrics to characterise the level of deliberation in discussion threads, and apply the model and metrics to a large Reddit dataset containing posts from 72 subreddits focused on different topics. We compare the level of deliberation that occurs on different subreddits, finding that subreddits that are based on geographical regions or focus on sports have the highest levels of deliberation. Analysis of the actor layer reveals several features consistent across all subreddits, such as small-world characteristics and similar numbers of highly active users.
\end{abstract}

\section{Introduction}
Online discussion platforms, such as Reddit and Quora, allow their users to discuss a range of topics, initiated via an original post, to which users can contribute through comment replies.\footnote{For reviews of how the different technological affordances of platforms affect how discussions unfold, see \cite{Medvedev_overview_Reddit} (Reddit), \cite{zimmer2014topology} (X, formerly Twitter), and \cite{wilson2012review_facebook} (Facebook).} Such platforms have become ubiquitous in modern society, and are increasingly central to social and political discourse \citep{noveck2009oneline_forum}. These platforms have the potential to foster opinion formation and consensus building in ways that traditional face-to-face discussions cannot due to the asynchronous nature of the interactions, the persistence of a comment chain recording the discussion, and the online aspect \citep{price2009citizens, howard2012socialmedia_review_paper}. These features allow for a diverse range of views from a greater number of users, including those who might be geographically separated, have busy schedules, or from minority and underrepresented groups~\citep{ulucc2010blogs_political}. Consequently, participants may be exposed to a broader range of views which may allow them to obtain a more comprehensive understanding of the issues at hand. However, online discussion platforms are increasingly implicated in information disorders such as the spread of false information, echo chambers and the facilitation of online harassment and hate speech~\citep{mendoza2023study_misinformation, kizilhan2016rise_polarisation, dahiya2021hates} which may contribute to societal problems such as polarisation and declining trust in institutions. For policymakers and researchers, it is therefore of substantial interest to be able to differentiate discussions that lead to positive as opposed to negative societal outcomes, and identify conditions that foster the former rather than the latter.


One approach to understanding the dynamics of online communication is to investigate user-to-user interactions. \cite{gomez2008statistical} built user-to-user interaction networks from discussions taking place on Slashdot, and characterised their structural properties, revealing both similarities (such as the small-world effect) and differences (such as the absence of degree assortativity) compared to traditional social networks. \cite{guerrero2018twitter_actor} examined retweeting, mentioning, and replying behaviours on Twitter during two political discussions in Spain from the 2015 and 2016 general elections, finding that Twitter users consistently exhibited homophilic behaviour during political campaigns. \cite{xiong2020understanding} utilised a dataset from government micro-blogs on Sina Weibo to construct directed user-to-user reply networks and investigated how network structures and attributes influence network formation. Their findings indicated that the user-to-user reply network is highly reciprocal and transitive, with users possessing greater social influence (defined as those with more followers or recognised as opinion leaders) being more likely to receive replies, but less likely to send replies.

A second approach involves examining the structure of the comments that form the discussion. Since most online platforms allow for users to contribute to an existing comment chain, with comments sequentially added in reply to those that already exist, the discussions are often represented using hierarchical networks~\citep{gonzalez2010structure}. Adopting this paradigm, \cite{medvedev2019modelling} developed a Hawke's process model for capturing how discussion threads grow over time, and fitted the model against Reddit data. 
In a similar vein, \cite{nishi2016twitterreply} constructed hierarchical networks based on reply interactions on Twitter, where the root was a tweet, and replies formed the branches. They found that the patterns of these networks were either long path-like, large star-like, or irregular. While such research provides insight into how the discussion unfolds over time, it provides limited ability to differentiate between those discussions which could contribute to consensus building instead of negative outcomes such as polarisation and increased hostility.

One framework that has the potential to address this is that of \textit{deliberation}, or \textit{deliberative democracy}~\citep{habermas1985remarks}. Theories of deliberative democracy were initially intended to address deficiencies in traditional democratic theory \citep{habermas1985remarks,dryzek1994australian}. These theories emphasise the importance of rational discourse and inclusive dialogue, suggesting that democratic legitimacy is best achieved through informed and respectful deliberation among citizens. It also suggests that not all discussions are equal, positing that \textit{deliberative} discussions are those which involve reasoned argumentation, critical engagement, and the inclusion of diverse perspectives. Such discussions can contribute to improving democratic institutions and processes, by allowing decision-making to follow after robust and thorough discussion. Subsequently, researchers have sought to refine the concept of deliberation. Ackerman and Fishkin, for instance, identified two key prerequisites for ideal deliberation: the involvement of a large number of participants and extensive exchange of views among them~\citep{ackerman2002deliberation}, referred to as \textit{representation} and \textit{argumentation}, respectively, for brevity.

In face-to-face discussions, these two prerequisites of deliberation can play contradictory roles in promoting deliberation. It was noted that increasing the number of participants raises the risk of the discussion being dominated by a small group of persuasive individuals, while more extensive argumentation can hinder spontaneous engagement in the debate~\citep{elster1998deliberative}. There has been growing interest in whether online discussion platforms are a potential solution to this contradiction~\citep{shane2004democracy}, with the ability to promote democracy. The Internet's role in reducing the transaction costs of participation, potentially fostering more robust political discussions and greater engagement from disenfranchised groups has been noted~\citep{romsdahl2005political}, as has the importance of institutions utilising Web 2.0 technologies to create democratic spaces that facilitate citizen involvement in policy-making~\citep{steibel2015designing}. 

\cite{gonzalez2010structure} proposed that the two aforementioned prerequisites of deliberation, viz. argumentation and representation, could be proxied by two metrics from a given hierarchical network representation of an online discussion thread. Namely, it was proposed that the number of nested layers, termed \textit{maximum depth} could capture argumentation, while the maximum number of comments at any layer, termed \textit{maximum width}, could capture representation.   
By comparing the maximum depth and maximum width between different posts on Slashdot, it was found that political discussions exhibited greater maximum depth and width than non-political discussions. The authors presented this as evidence of the construct validity of their network measure of deliberation (high maximum depth and width of discussion trees) since one would expect deliberation to occurring in political, compared with non-political, discussions.

Building on this work, \cite{ackland2023reciprocal} incorporated reciprocal communication and political partisanship of users into measures of deliberative potential, and applied this to Twitter data. Using these enhanced metric, they found that the discussions on Twitter associated with the first debate of the 2020 US presidential election did not exhibit wide-scale deliberation. The effect of the sentiment of the comments on the deliberative potential of discussions has also been examined~\citep{Argon_2017_sentiment}. It was found that comments negatively aligned with the original post/comment were more likely to trigger discussion cascades, suggesting that negative comments may contribute to the argumentation by provoking more substantive exchanges.

\subsection{Main contributions of this paper}
While there is extensive literature on the analysis of the structure of user-to-user networks in online discussion, and the structure of discussion threads and its connection with deliberation, research integrating these two aspects remains limited. Furthermore, there remain limited methods to quantify the ``level of deliberation'' for online discussions; expanding such methods would allow for better comparisons of the deliberative potential of different sets of online discussions (where the sets may be discussions on different topics, or from different groups of users). Our study aims to address these two gaps.

We propose a two-layer network model to describe discussions in online platforms, comprising a \textit{discussion layer} and an \textit{actor layer}. The former is inspired by Gonzalez-Bailon \textit{et al.}'s use of hierarchical networks to capture the structure of the comments, while the latter captures user-to-user interactions. Inter-layer links capture the ownership of comments by specific users. We apply this two-layer model to Reddit. In particular, we collected a large dataset of posts from 72 highly active subreddits (sub-forums that focus on specific topics) across a range of categories such as sports, politics, and region-based subreddits. We applied the two-layer model to each subreddit (so the discussion layer captures all the posts from a single subreddit, and the actor layer includes all users who contributed to said subreddit).

Moving beyond the maximum width-maximum depth metric, we introduce two additional metrics: average width-average depth, and dyadic conversation count-average dyadic conversation length. These two new metrics give additional insights into the deliberative potential of a discussion, particularly when the hierarchical tree shows significant asymmetry. We propose a quantitative method to measure and compare the level of deliberation across different groups of posts and apply these additional metrics and method to the Reddit data we collected. 

We find that subreddits focused on sports or a particular geographical region consistently had high levels of deliberation across all three metrics, higher than political subreddits. In contrast, other subreddits ranked highly in one metric but not in another, suggesting that the discussions occurring on such subreddits had fundamentally different structural properties. For example, subreddits focused on humorous content showed high levels of deliberation in the maximum width-maximum depth metric but demonstrated significantly lower levels in the average width-average depth metric. We also analyse the structural characteristics of actor layer networks within different subreddits and identify several consistent features: the networks are highly sparse, follow a power-law degree distribution, exhibit small-world characteristics, have a similar number of highly active users, and show moderate degree assortativity. Interestingly, despite the difference in subscriber counts between different subreddits, the number of users who contributed frequently was far more similar across all subreddits. 

In summary, this paper provides a methodological advance in the analysis of deliberation in online discussion platforms via the two-layer network model and associated metrics of deliberation. We also provide specific insight on Reddit, including contrasting characteristics between subreddits, by applying the proposed methodology to a large dataset.

The rest of the paper is structured as follows. In Section~\ref{Sec:two_layer_model}, we give an overview of Reddit, introduce the two-layer network model and explain the Reddit dataset. We also introduce the associated metrics for deliberation, comparing the existing metrics with our newly proposed ones. Section~\ref{sec:subreddit} examines the metrics with respect to different subreddits, while Section~\ref{sec:actor_layer} investigates the actor layer network structures. Conclusions and future work are drawn in Section~\ref{sec:conclusions}.

\section{Methodology} \label{Sec:two_layer_model}

In this section, we first give a basic description of Reddit. We then present a two-layer network model which is used to represent discussions in online social media platforms, and associated metrics for measuring the level of deliberation in a post or discussion thread. While this paper focuses on discussions on the Reddit platform, our model is general enough such that it can be adapted to a range of different platforms that involve posting of sequential comments by different users.

\subsection{Overview of the Reddit platform}
Reddit is an American discussion platform with substantial international usage, where the vast majority of users participate anonymously. Posts on Reddit are organised into user-generated groups known as subreddits. Each subreddit is dedicated to a specific topic or theme, and users can subscribe to these subreddits to receive content related to their interests, or browse a `Front Page' which presents the most popular posts from different subreddits, where popularity can be filtered by country or globally. This model is different from traditional social networks like Facebook or Twitter, where users typically follow other individual users. The subreddit-based organisation of Reddit fosters relatively independent groups of users focused around specific topics. Each subreddit operates like a mini-society with its own rules and culture, moderated by other users.

In the Reddit environment, a discussion starts with a single user creating an original post (for simplicity, in this paper, we consider the original post as a comment and refer to it as the original comment). Then, other users can contribute by adding subsequent comments to the original comment, sparking further discussion within those comments. This nested structure allows users to engage with existing comments or the original comment itself, generating new comments that in turn invite more interaction. A Reddit discussion can be portrayed as a directed radial tree. Within this tree representation, nodes represent comments within the post, while a directed edge represents reply relations, so that the head node is a comment posted in reply to the tail node comment. The original comment, acting as the source node, is positioned at the first level. Comments directly responding to the original comment occupy the second level, while subsequent comments, forming additional nested levels, unfold to create new branches or levels in the evolving tree structure. As illustrated in the lower part of Fig.~\ref{fig:two_layer_model} (in the blue background frames), two examples of directed radial trees depict distinct Reddit discussions. Each level of the trees is distinguished by a unique color scheme: nodes at levels 1, 2, 3, and 4 are depicted in red, black, green, and yellow, respectively.

\subsection{Description of two-layer model}

\begin{figure} 
	\centering
	\includegraphics[width=0.9\linewidth]{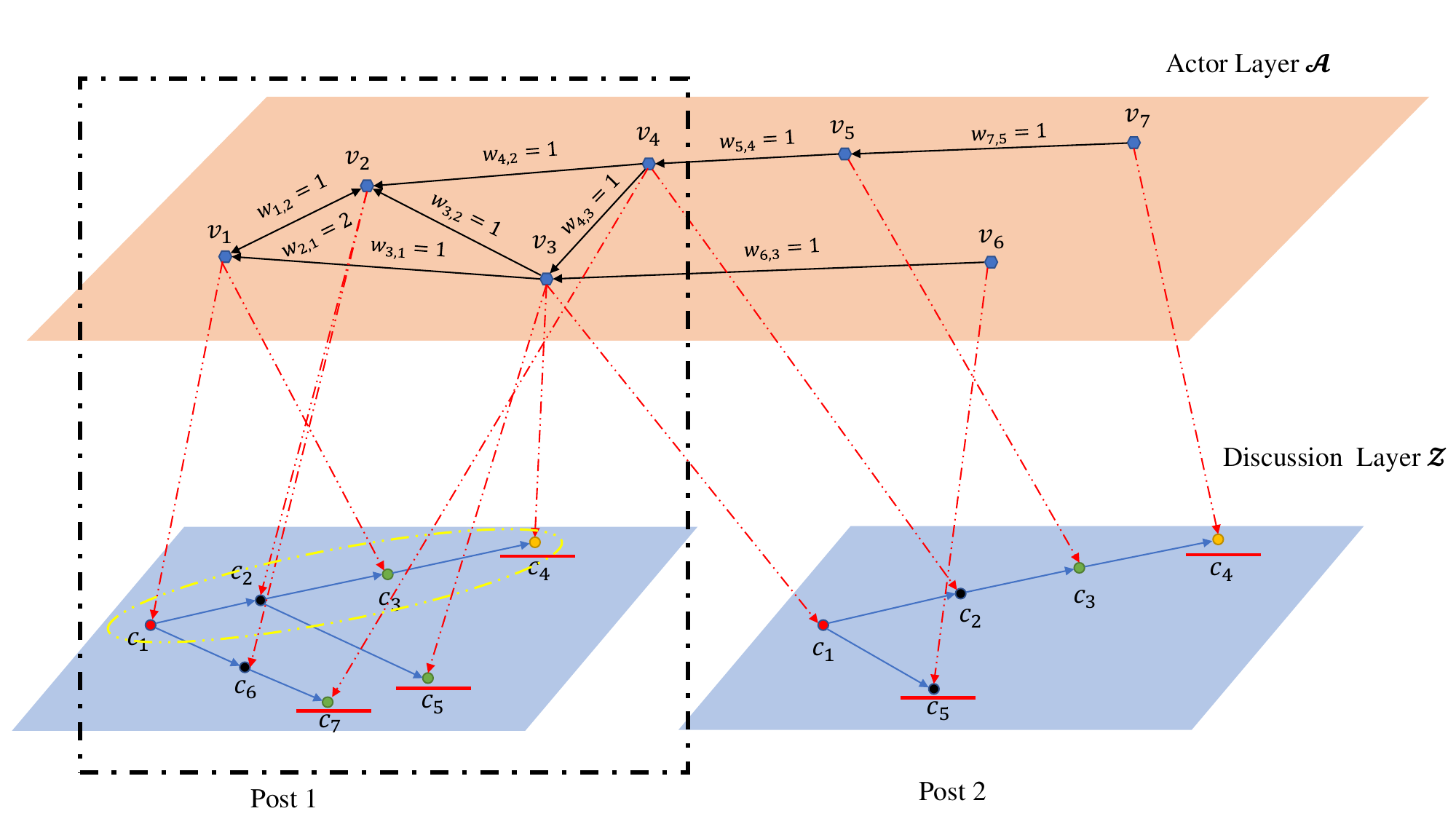}
	\caption{Initially, this two-layer network is based on Post 1, enclosed within a black dashed box. The discussion layer is structured as a radial tree, where nodes represent comments. Edges between these nodes, depicted by blue arrows, indicate reply interactions. Sink nodes are highlighted with red outlines. The actor layer, on the other hand, consists of four users contributing to Post 1, resulting in four distinct nodes. Edges between these nodes, represented as black arrows, signify one user replying to the other's comment. Meanwhile, the edges between the actor layer and the discussion layer, shown as red dashed arrows, illustrate the users responsible for these comments. The dyadic conversation is highlighted within the yellow dashed oval in the discussion layer. Subsequently, the model expand to Post 2, where the activity layer and the actor layer includes five and three additional nodes, respectively. The edges and weights within the actor layer network is also expanded accordingly. 
		 \vspace{-11pt}} 
	\label{fig:two_layer_model}
\end{figure}

Our two-layer network model is represented by a two-layer graph $\mathcal{G}=(\mathcal{V}, \mathcal{E}, \mathcal{W}, \mathcal{C}, \mathcal{F}, \mathcal{H})$, comprising of an actor layer $\mathcal{A} = (\mathcal{V}, \mathcal{E}, \mathcal{W})$ and a discussion layer $\mathcal{Z} = (\mathcal{C}, \mathcal{F})$. For simplicity, we will elucidate the model constructed based on a single post (depicted in Fig.~\ref{fig:two_layer_model} within the black dashed box). In order to analyse posts from entire subreddits, the model is expanded below to encompass multiple posts (as shown in Fig.\ref{fig:two_layer_model} depicting two posts). 

The discussion layer contains the node set $\mathcal{C} = \{c_1, c_2, \hdots, c_r\}$, which represents the comments of a single discussion on Reddit initiated from a post. The node $c_1$ represents the original comment, while $c_k$ for $k \geq 2$ represents subsequent comments within the discussion. An edge $(c_i, c_j)$ in the edge set $\mathcal{F}$, going from $c_i$ to $c_j$, represents that comment $c_j$ is a reply to comment $c_i$. Because of the sequential nature of the comments (one can only write a comment in reply to an existing comment), it follows that $i < j$, and thus the graph on the discussion layer is a directed tree graph, where the original comment $c_1$ is the only source node, and the node $c_j$ such that $(c_j,c_k) \notin \mathcal{F}$ for any $c_k \in \mathcal{C}$ is a sink node. 

The actor layer captures the users who contributed comments to the post currently being modelled, with $\mathcal{V} = \{v_1, v_2, \hdots, v_n\}$ being the $n$ users who contributed to the post. The edge set $\mathcal{H}\subseteq \mathcal{V} \times \mathcal{C}$ connects nodes from the actor layer to the discussion layer, capturing which user contributed which comment. Namely, $(v_i, c_k)\in\mathcal {H}$ indicates that comment $c_k\in\mathcal C$ was made by user $v_i \in \mathcal V$. It is worth noting that each comment can only be attributed to one user. Therefore, if $(v_i, c_{u})\in \mathcal{H}$ then $(v_j, c_{u}) \notin \mathcal{H} \mid \forall v_j\in\mathcal{V}\setminus v_i$.

The edge set $\mathcal{E} \subseteq \mathcal V\times \mathcal{V}$ captures the reply relations between users (also termed as actors). Namely, an edge $(v_i, v_j)$, which goes from $v_i$ to $v_j$, indicates that user $v_i$ has posted a comment $c_k$ in reply to a comment $c_\ell$ that belongs to user $v_j$. In other words, the edge $(v_i, v_j)$ exists if and only if there exists an edge $(c_\ell, c_k)$ in $\mathcal Z$ such that $(v_i, c_k)$ and $(v_j, c_\ell)$ are both in $\mathcal H$. Since users can make multiple comments within a single post, we use an edge weight $w_{ij} \geq 0$ to record the number of times user $v_i$ has replied to $v_j$. In particular, $w_{ij} = 0$ if and only if $(v_i, v_j) \notin \mathcal E$. Otherwise, \[w_{ij} = \vert \{(c_\ell, c_k) : (v_i, c_k) \text{ and } (v_j, c_\ell) \in \mathcal H\} \vert.\]
We collect all the edge weights $w_{ij}$ into the weight set $\mathcal W$.

\graphicspath{{figures/}}

\subsection{Metrics for the two-layer model}

We now present six network-based metrics for the two-layer model: maximum width, maximum depth, average width, average depth, average dyadic conversation length, and dyadic conversation count (Table~\ref{tab:metric discription} provides a summary of these metrics). 

To begin, we define the depth of node $c_k$ as the number of edges in the path from the source node $c_1$ (the original comment) to node $c_k$, and denote this by $D_k\in \mathbb N_{\geq 0}$ (the depth of the source node is $c_1=0$). The maximum depth of the discussion tree, $\hat D$, is the maximum depth across all nodes in $\mathcal C$: \[\hat D = \max_{c_k\in\mathcal C} D_k.\]
Note that necessarily, the node with the maximum depth must be a sink node. The average depth of the discussion tree, $\bar D$, is the average depth of all the sink nodes in the discussion tree. More formally, if $\mathcal C'\subset \mathcal C$ denotes the set of sink nodes, then \[\bar D = \frac{1}{\vert \mathcal C'\vert} \sum_{c_k\in\mathcal C'} D_k.\]

Meanwhile, the width of the discussion tree at depth $q\in\mathbb N$, denoted by $W_q$, is equal to the number of nodes with depth $q$. That is, $W_q = \vert {c_k \in \mathcal C : D_k = q} \vert$. The maximum width of the discussion tree, $\hat W$, is the maximum width at any depth of the tree, i.e. \[\hat W = \max_{q\in \mathbb N} W_q.\] 
The average width of the discussion tree, $\bar W$, is the arithmetic mean of the widths at each depth, i.e., \[\bar W = \frac{1}{\hat D} \sum_{q=1}^{\hat D} W_q.\]

To define the last two metrics, we first need to define the concept of a dyadic conversation. Due to the sequential commenting nature of Reddit, and many other online platforms, we define a dyadic conversation as a sequence of comments involving just two users who reply to each other. That is, a dyadic conversation $y$ is a sequence of $m\geq 3$ nodes in the discussion layer (a branch of the tree) where the inter-layer edges connect to just two users, say $v_i$ and $v_j$, in alternating fashion. That is, if the dyadic conversation included a sequence of nodes $c_p, c_q, c_r$, then $(v_i, c_p)$, $(v_j, c_q)$, and $(v_i, c_r)$ belong to $\mathcal{H}$.\footnote{Note that it is possible for there to be more than one dyadic conversation within a given path from source node to sink node.} The length of a dyadic $y$ is defined as $m$. Within the discussion tree, there may be multiple dyadic conversations between multiple pairs of users. Let $\mathcal{Y} = \{y_1, y_2, \hdots, y_k\}$ be all the dyadic conversations in the discussion tree, with conversation $y_i$ having length $m_i$. We define $\bar y = \vert \mathcal{Y}\vert$ as dyadic conversation count, and $\bar x = \frac{1}{\bar y} \sum_{i=1}^k m_{i}$ be average dyadic conversation length.

\begin{table}[!ht]
    \centering
    \begin{tabular}{lll}
    \hline
        Metric & Notation & Discription \\ \hline
        Maximum depth & $\hat{D}$ & The maximum depth across all nodes in the discussion tree \\ 
        Average depth & $\bar{D}$ & The arithmetic mean of the depth of all sink nodes \\ 
        Maximum width & $\hat{W}$ & The maximum width at any depth of the discussion tree \\ 
        Average width & $\bar{W}$ & The arithmetic mean of the width at each depth \\ 
        Dyadic conversation count & $\bar{y}$ & The number of reciprocal conversations within the post \\ 
        Average dyadic conversation length & $\bar{x}$ & 	The arithmetic mean of the length of all reciprocal conversations within the post \\ \hline
    \end{tabular}
    \caption{Summary of discussion tree metrics}
    \label{tab:metric discription}
\end{table} 

\subsection{Data collection and the two-layer model for a single subreddit}

We used the website \href{https://subredditstats.com/}{https://subredditstats.com/} to identify the top 70 most frequently commented subreddits and the 20 subreddits with the highest comment growth for our data collection.\footnote{The website \href{https://subredditstats.com/}{https://subredditstats.com/} ceased updating following Reddit's implementation of new API rate limit rules in July 2023, which occurred after we completed data collection.} We then used the R package vosonSML \citep{vosonSML} to collect the Reddit data. The data collection process involved two primary steps. Firstly, every Monday, we gathered post URLs from these top 90 subreddits, encompassing all posts created within the preceding week. Subsequently, we refined the collected posts by excluding those created less than 24 hours prior, after noting that approximately 98\% of comments typically arise within this initial time-frame. In other words, this refinement ensured that we collected posts for which the discussion has effectively concluded. We further filtered the posts, including those with a comment count ranging from 5 to 1400. The lower threshold stems from our research focus on deliberation, as posts with fewer than 5 comments are not relevant. The upper threshold aligns with the limitations imposed by the Reddit API, as we are constrained to collect a maximum of 1400 comments per post. Given that less than 0.1\% of posts exceed the upper threshold, we believe that this refinement process had minimal impact on our research findings. We repeated this data collection process over the course of 8 weeks, from July 18, 2023, to September 12, 2023. 

Finally, we narrowed down the initial 90 subreddits to 72 by applying the following criteria. We selected subreddits that had accumulated a minimum of 1,500 posts over the 8-week period, along with a minimum of 1,500 unique users who contributed comments at a frequency of more than two posts per week. The list of the final 72 subreddits is provided in Table~\ref{tab:subreddit_group}. Altogether, we gathered a total of 20,047,470 comments contained in 218,892 posts.  
 
\begin{table}[!ht]
    \centering
    \resizebox{\textwidth}{!}{
        \begin{tabular}{ l l l l l l l }
\hline
        \textbf{SPORTS} & \textbf{REGIONAL} & \textbf{POLITICS} & \textbf{GAMING} & \textbf{FUNNY} & \textbf{KNOWLEDGE/OPINION SHARING} & \textbf{ENTERTAINMENT} \\ \hline
        8 baseball & \textbf{General} & 43 neoliberal & \textbf{General Games} & 1 AITAH & 41 namenerds & 9 books \\ 
        13 CFB & 6 australia & 51 politics & 28 Games & 2 amiugly & 67 UFOs & 40 movies \\ 
        16 Cricket & 11 canada & 72 worldnews & 29 gaming & 62 TooAfraidToAsk & 61 Tinder & 58 television \\ 
        24 formula1 & 23 europe & 49 PoliticalCompassMemes & 48 pcgaming & 4 AskMen & 34 LifeProTips & 10 boxoffice \\ 
        30 golf & 65 Turkey & ~ & 53 PS5 & 25 FragReddit & 17 CryptoCurrency & 35 LoveIslandTV \\ 
        31 hockey & 68 unitedkingdom & ~ & \textbf{Specific Games} & 7 aww & 70 wallstreetbets & 38 marvelstudios \\ 
        39 MMA & \textbf{With restriction} & ~ & 20 DestinyTheGame & 15 CrazyFuckingVideos & 66 TwoXChromosomes & 45 OnePiece \\ 
        42 nba & 5 AskUK & ~ & 21 DnD & 18 Damnthatsinteresting & 3 antiwork & 46 OnePiecePowerScaling \\ 
        44 nfl & 12 CasualUK & ~ & 22 Eldenring & 36 MadeMeSmile & 64 TrueUnpopularOpinion & 57 TaylorSwift \\ 
        54 soccer & 32 KGBTR & ~ & 27 GachaClub & 59 therewasanattempt & 69 unpopularopinion & 63 travisscott \\ 
        56 SquaredCircle & ~ & ~ & 33 leagueoflegends & 37 MapPorn & ~ & ~ \\ 
        ~ & ~ & ~ & 47 Overwatch & 60 TikTokCringe & ~ & ~ \\ 
        ~ & ~ & ~ & 55 SpidermanPS4 & 71 WhitePeopleTwitter & ~ & ~ \\ 
        ~ & ~ & ~ & ~ & 19 dankmemes & ~ & ~ \\ 
        ~ & ~ & ~ & ~ & 26 funny & ~ & ~ \\ 
        ~ & ~ & ~ & ~ & 52 ProgrammerHumor & ~ & ~ \\ 
        ~ & ~ & ~ & ~ & 50 PoliticalHumor & ~ & ~ \\ 
        ~ & ~ & ~ & ~ & 14 conspiracy & ~ & ~ \\ \hline
        \end{tabular}
    }
    \caption{Index and categorisation of subreddits and their sub-categories. The index is organised alphabetically based on the subreddit names.}
    \label{tab:subreddit_group}
\end{table}

In this paper, we are interested in exploring differences between subreddits in terms of levels of deliberation. Thus, we apply our two-layer model to each subreddit separately. Specifically, the discussion layer is constructed from \textit{all} the posts collected for that subreddit (after following the filtering process described above). Thus, the discussion layer contains thousands of discussion trees, which are disconnected from each other. We compute the six metrics mentioned above,
for each post separately. The actor layer thus contains all users who contributed comments to the subreddit during the 8 week data collection period, with many actors contributing to multiple different posts over time. In doing so, we create a two-layer network that captures all the users and comments for a given subreddit. Our results, presented below, will focus on comparing the metrics of deliberation, as computed using the two-layer networks, between different subreddits.

\subsection{Existing metrics for deliberation}

We build on \cite{gonzalez2010structure}'s foundational work on network metrics for deliberation in online discussion forums. Using Slashdot data collected between 2005 and 2006, \cite{gonzalez2010structure} proposed maximum depth and maximum width as proxies for the level of \textit{argumentation} and \textit{representation}, respectively, which are the two key prerequisites for a discussion to be considered deliberative. In other words, the high levels of deliberation were posited to occur when the discussion tree had high maximum width and maximum depth. The authors computed $\hat W$ (maximum width) and $\hat D$ (maximum depth) for each post and defined four different types of posts, depending on where the post was situated on a two-dimensional coordinate system of maximum width and maximum depth. Type I networks were expected to have high levels of deliberation, while Type II and Type IV networks only had high maximum depth and high maximum width, respectively. Type III networks were the least deliberative, with a low maximum width and maximum depth. By differentiating between political posts and non-political posts, it was found that political posts exhibit high levels of deliberation. While \cite{gonzalez2010structure} introduced a major conceptual innovation by using width and depth to estimate levels of deliberation and then used these metrics to illustrate that political posts had higher levels of deliberation than non-political posts, there are several limitations.

The first limitation is best seen when comparing two example posts we collected from Reddit\footnote{Post 1:~\url{https://www.reddit.com/r/travisscott/comments/16df2o8/rodeo\_but\_hes\_looking\_directly\_at\_you/}\\
Post 2:~\url{https://www.reddit.com/r/Games/comments/16esm4n/dying\_light\_2\_had\_200\_pages\_of\_cut\_content/}}, with the discussion tree presented in Fig.~\ref{fig:two_post_comparison}. Specifically, Post 1 and Post 2 exhibit identical maximum depth and maximum width values. Despite these metrics being equivalent, Post 2 has a greater number of comments and by visual inspection, it is not unreasonable to conclude that it has a more vigorous intensity of argumentation compared to Post 1. This discrepancy arises from an inherent assumption when using the maximum depth and width as metrics for deliberation: the structure of the discussion tree, namely the length of each branch, is homogeneous. However, this assumption is clearly not appropriate in the case of Post 1, where branches are highly heterogeneous in length. Consequently, in such networks, the use of maximum width and maximum depth metrics may potentially overestimate the degree of deliberation.

The second limitation is that the dimensions of depth and width only concern the discussion layer and ignore information from the actor layer, such as the characteristics and identities of the users. This limitation may result in an incomplete understanding of the unfolding conversation. For instance, consider the longest branch in Fig.~\ref{sfig:post_1_thread}.
The same branch could arise when every comment is made by independent users (i.e., each user makes only one comment without further replies) or when just two users engage in a prolonged discussion, consistently responding to each other and extending the chain. Evidently, these two scenarios involve different discussion dynamics, and this information can be identified using the two-layer network and the associated dyadic conversation count and length metrics, which we will explain below.

The final issue, not restricted to the work of \cite{gonzalez2010structure}, is the lack of a grounded approach to quantify the difference in deliberation between multiple sets/collections of posts. For instance, \cite{gonzalez2010structure} compares the centroid value of the maximum width and maximum depth between political posts and non-political posts from Slashdot data; since the centroid coordinates are greater on both axes for political posts, it is concluded that political posts on average are more deliberative than non-political posts. However, the centroid coordinates involve two dimensions (depth and width), and it can thus be difficult to compare two or more sets of posts if some sets excel along one dimension while others excel alone the other. It is therefore not as straightforward to quantitatively rank deliberation levels across multiple sets.



\subsection{Revised metrics for deliberation}
\begin{figure*}	
	\centering
	\subfigure [] {
		\includegraphics[width=0.45\linewidth]{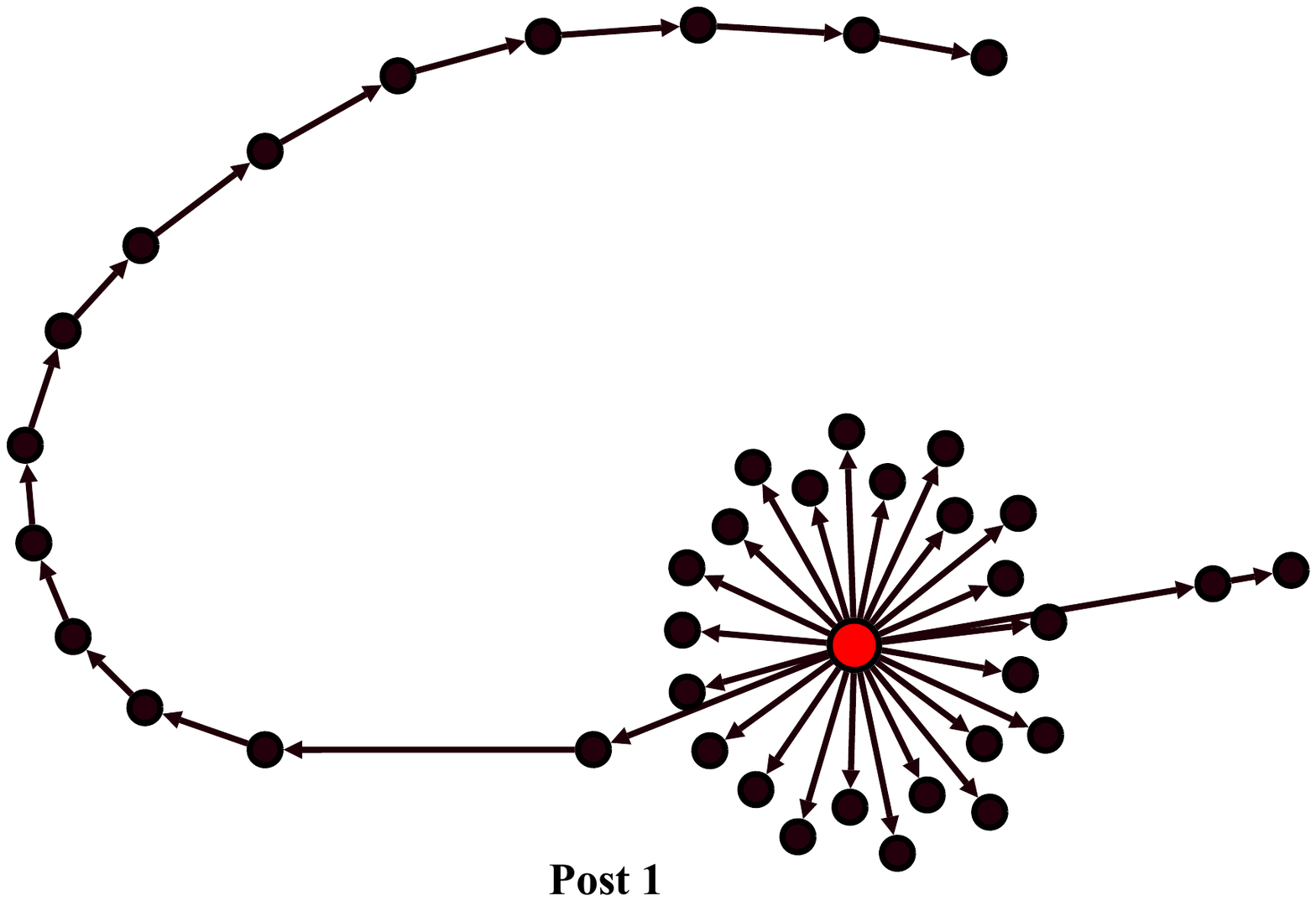} 
		\label{sfig:post_1_thread}
  } \subfigure [] {
		\includegraphics[width=0.45\linewidth]{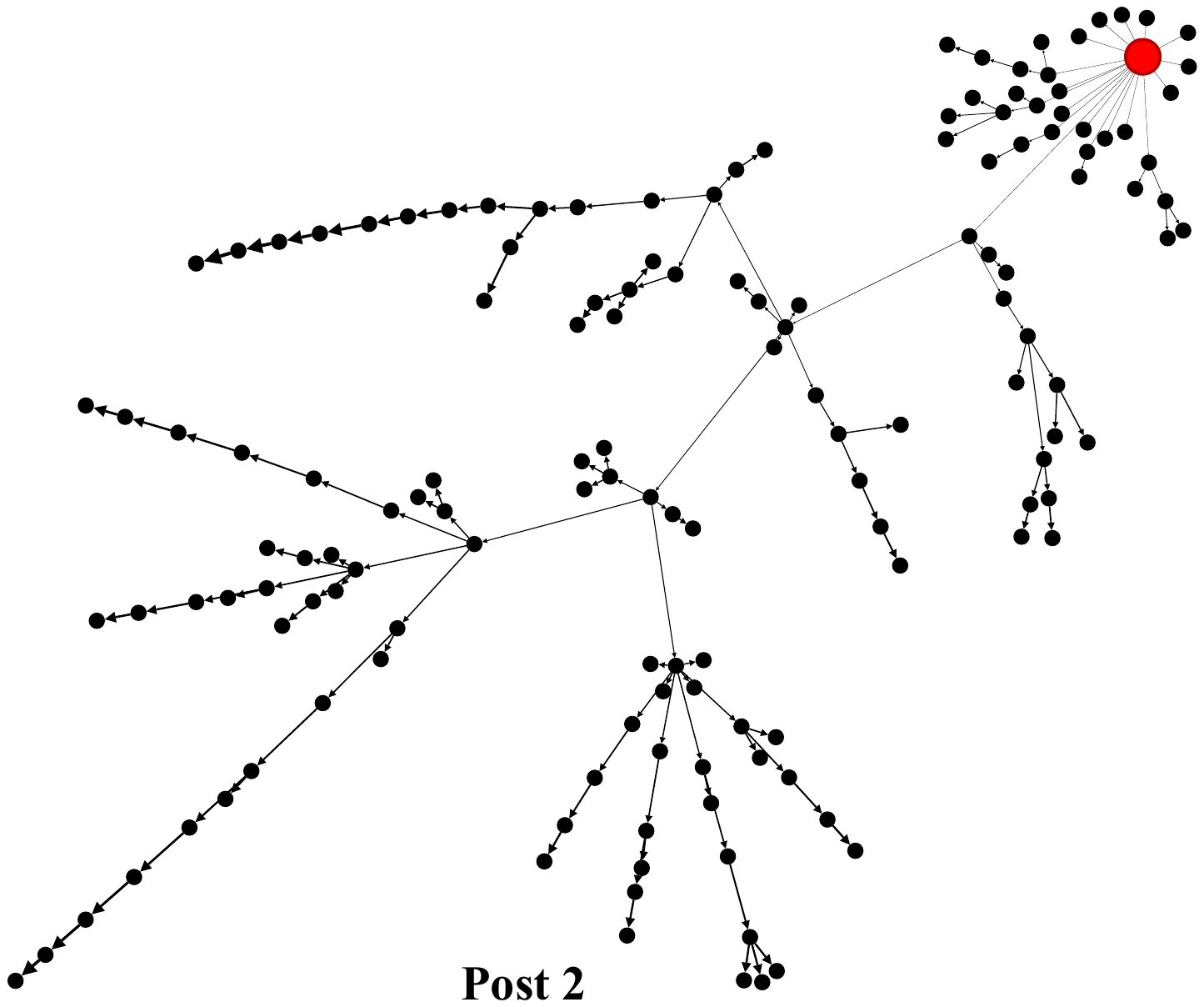}
		\label{sfig:post_2_thread}
             }
\caption{Comparison of the thread structure of two post. The maximum depth and width for these two posts are both 14 and 25, respectively. The average depth for Post 1 and Post 2 is 1.56 and 4.89, respectively. The average width for Post 1 and Post 2 is 2.78 and 10.4, respectively. The original posts are red nodes.}  
 \label{fig:two_post_comparison}
\end{figure*}

We propose to address the first limitation by introducing average width and average depth metrics, which aim to reduce the impact of outlier branches in the discussion layer with exceptionally high depth. In the example shown in Fig.~\ref{fig:two_post_comparison}, Post 2 has average depth and width values of 4.89 and 10.4, respectively, which are significantly higher than those of Post 1 (1.56 and 2.78). This disparity reflects the difference in the degree of deliberation between these two posts. The second limitation is addressed by introducing dyadic conversation metrics, which are possible only by using the two-layer model, as dyadic conversations are identified by examining both layers of the network together.

The third limitation is addressed by introducing a method for drawing the boundaries using a threshold parameter, and then demonstrating that our conclusions are robust with respect to the parameter value. First, we plot every post from all 72 subreddits onto a two-dimensional scatter plot, with $X-Y$ coordinate pairs being i) maximum width-maximum depth, ii) average width-average depth, and iii) dyadic conversation count-average dyadic conversation length. See Fig.~\ref{plot_transfer} for an example where a subset of posts from one subreddit is plotted onto two of the coordinate pairs. We use the following criteria for creating the boundary lines demarcating the four quadrants.
\begin{enumerate}
	\item Quadrant~$\text{I}$ contains $10\%$ of all the posts collected across the 72 subreddits.
	\item Quadrants~$\text{II}$ and $\text{IV}$ contain a similar proportion of posts (i.e., 18.0\% and 18.2\%, respectively).
\end{enumerate}
For a particular subreddit, we then define $P_{10}^a$ as the fraction of posts from that subreddit that are in Quadrant~$\text{I}$, when using the average width-average depth coordinate system. We define $P_{10}^m$ and $P_{10}^d$ as the fraction of posts from that subreddit that are in Quadrant~$\text{I}$, when using the maximum width-maximum depth coordinate system, and dyadic conversation count-average dyadic conversation length coordinate system, respectively. We compare the level of deliberation between different subreddits by comparing the metrics $P_{10}^a$, $P_{10}^m$ and $P_{10}^d$, computed for the different subreddits, with higher values of these metrics corresponding to that subreddit having higher levels of deliberation among its posts. Thus, we are able to obtain three separate rankings of the 72 subreddits in terms of level of deliberation, using the three metrics of $P_{10}^a$, $P_{10}^m$ and $P_{10}^d$. 

\graphicspath{{figures/average_necessity/}}


\subsection{Robustness of the deliberation metric}

To test the robustness of the criteria for computing $P_{10}^a$, we repeated the same process, but identified Quadrant~$\text{I}$ by requiring that it contains $X\%$ of all posts collected across the 72 subreddits, whereas previously $X = 10$. We repeated the process for $X = 5, 20, 30$ and $50$, in order to additionally obtain metrics $P_{5}^a$, $P_{20}^a$, $P_{30}^a$ and $P_{50}^a$, respectively, for the 72 subreddits. Analysis using Spearman correlation shows that there is a high correlation between subreddit rankings when using the metrics $P_{5}^a$, $P_{10}^a$, $P_{20}^a$, $P_{30}^a$ and $P_{50}^a$, as detailed in Table~\ref{table: Spearman correlation of ranking}. In other words, the relative ranking of deliberation between subreddits is robust to the choice of $X$, where $X$ is used to define the boundary lines for Quadrant~$\text{I}$. For the rest of the paper, we focus on $X=10\%$, i.e., we use $P^a_{10}$.

\begin{table}[!ht]
    \centering
    \begin{tabular}{l l l l l l }
    \hline
        ~ & Rank by $P_{5}^a$ & Rank by $P_{10}^a$ & Rank by $P_{20}^a$ & Rank by $P_{30}^a$ & Rank by $P_{50}^a$ \\ \hline
        Rank by $P_{5}^a$  & 1 & 0.988 & 0.939 & 0.896 & 0.832 \\ 
        Rank by $P_{10}^a$ & 0.988 & 1 & 0.972 & 0.937 & 0.883 \\ 
        Rank by $P_{20}^a$& 0.939 & 0.972 & 1 & 0.989 & 0.956 \\ 
        Rank by $P_{30}^a$ & 0.896 & 0.937 & 0.989 & 1 & 0.983 \\ 
        Rank by$P_{50}^a$ & 0.832 & 0.883 & 0.956 & 0.983 & 1 \\ \hline
    \end{tabular}
      \caption{Spearman correlation of subreddit rankings based on $P_{5}^a$, $P_{10}^a$, $P_{20}^a$, $P_{30}^a$, $P_{50}^a$}
  \label{table: Spearman correlation of ranking}
\end{table}

\begin{figure*}	
	\centering
	\subfigure [] {
		\includegraphics[width=0.45\linewidth]{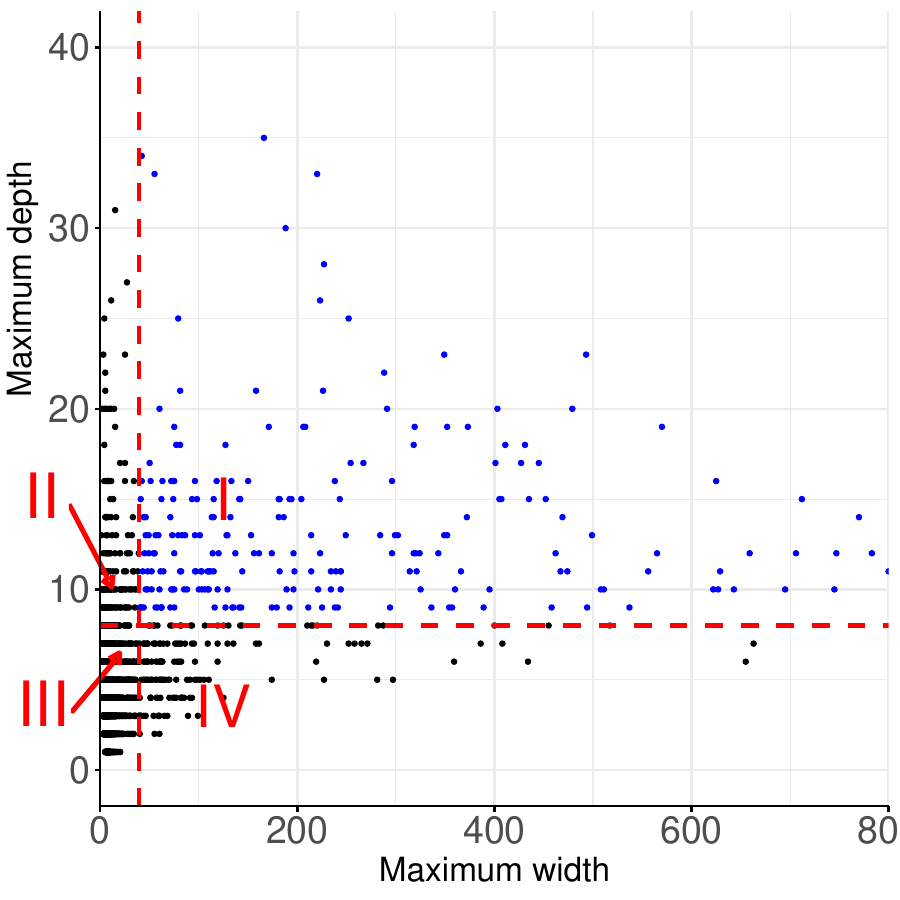} 
		\label{sfig:maximum_plot}
	} \subfigure [] {
		\includegraphics[width=0.45\linewidth]{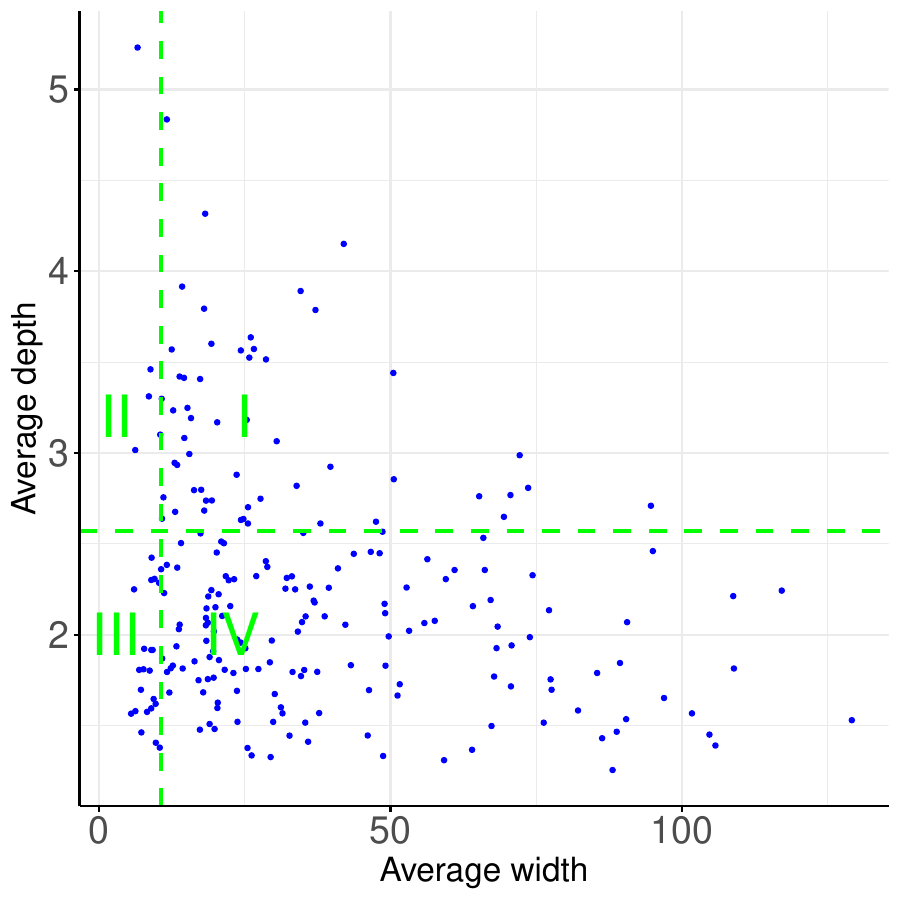}
		\label{sfig:max_ave_transfer}}
\caption{Comparison between the `maximum width-maximum depth' metric and the `average width-average depth' metric. In (a), all posts within the `AITAH' subreddit are plotted on the `maximum width-maximum depth' coordinate system. The red lines divide the coordinate system into four quadrants, with Quadrant \uppercase\expandafter{\romannumeral1} containing 10\% of posts across all 72 subreddits. In (b), the posts located in Quadrant \uppercase\expandafter{\romannumeral1} in (a) are selected and plotted on `average width-average depth' coordinate system. The green lines divide the coordinate system into four quadrants, with Quadrant \uppercase\expandafter{\romannumeral1} containing 10\% of posts across all 72 subreddits.}\label{plot_transfer}
\end{figure*}

To highlight the usefulness of the `average width-average depth' metric, we use the subreddit `AITAH' as an illustrative case. As depicted in Fig.~\ref{sfig:maximum_plot}, we first map all posts within the `AITAH' subreddit onto the `maximum width-maximum depth' coordinate system. Subsequently, we select the posts located in Quadrant~$\text{I}$, as defined above, and plotted these selected posts onto the `average width-average depth' coordinate system, as depicted in Fig.~\ref{sfig:max_ave_transfer}. Notably, only 23.9\% of the selected posts remain in Quadrant~$\text{I}$ of the new coordinate system. This suggests that a significant number of posts, initially showcasing high maximum width and depth, do not maintain high average width and depth. 

\begin{figure*}	
	\centering
	\subfigure [] {
		\includegraphics[width=0.45\linewidth]{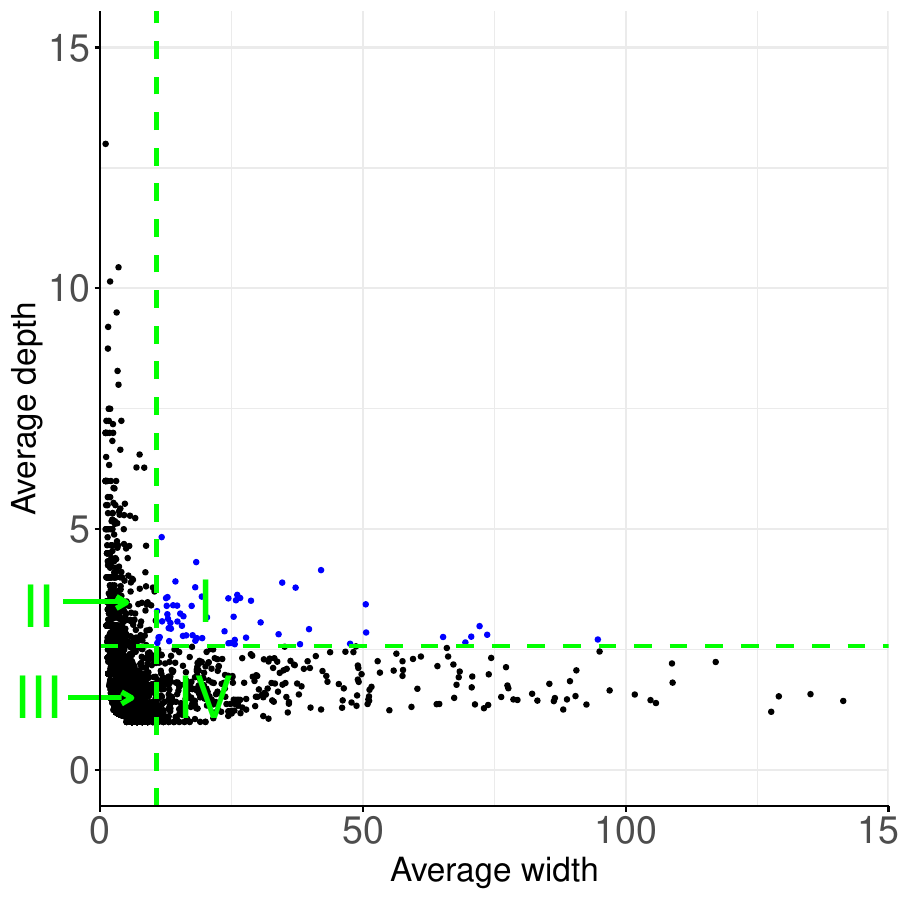} 
		\label{sfig:AITAH_ave_PLOT}
  } \subfigure [] {
		\includegraphics[width=0.45\linewidth]{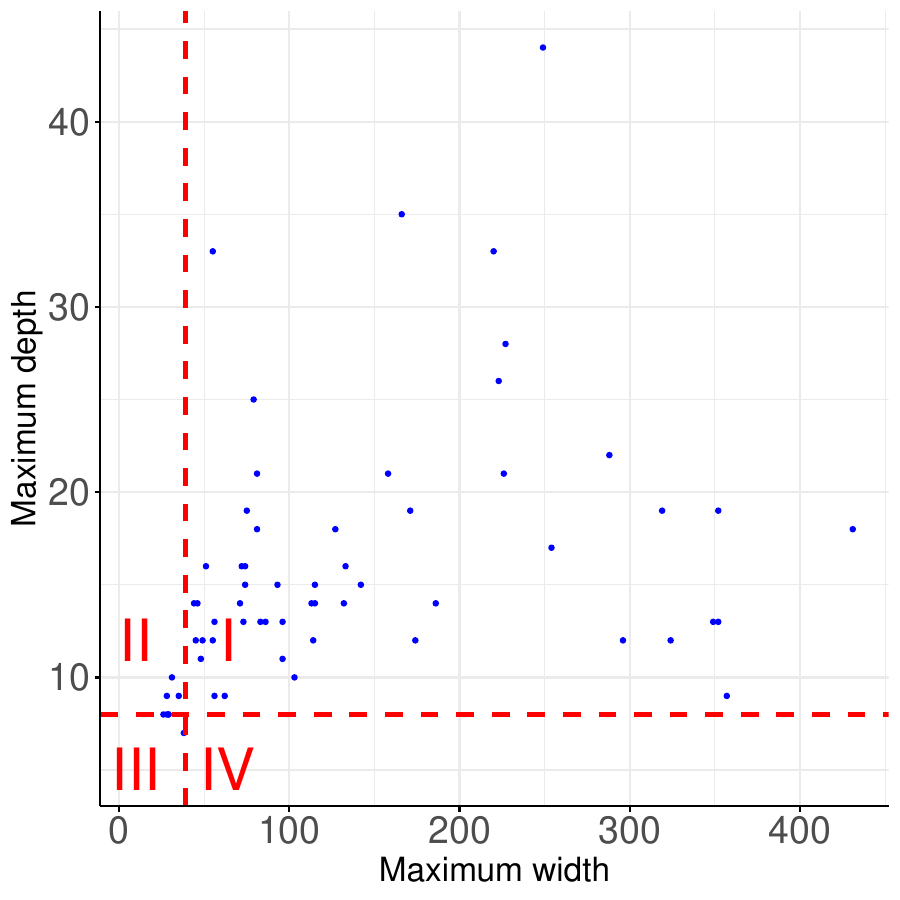}
		\label{sfig:AITAH_ave_max}
             }
\caption{Reversing the process in Fig.~\ref{plot_transfer}. In (a), all posts within the `AITAH' subreddit are plotted on the `average width-average depth' coordinate system. In (b), the posts located in Quadrant \uppercase\expandafter{\romannumeral1} in (a) are selected and plotted on the `maximum width-maximum depth' coordinate system. 
}
 \label{1}
\end{figure*}

Conversely, we reverse the process, with results shown in Fig.~\ref{1}. We first plot all posts within the `AITAH' subreddit onto the `average width-average depth' coordinate system. Subsequently, we select posts situated in Quadrant~$\text{I}$ and plot them in the `maximum width-maximum depth' coordinate system. Notably, most of the selected posts continue to be situated in Quadrant~$\text{I}$ in the `maximum width-maximum depth' coordinate system. This indicates that posts with high average width and depth are more likely to also possess high maximum width and depth, but the converse is not necessarily true. It also underscores the importance of expanding the measurement of deliberation to also include average width and average depth, rather than relying solely on maximum width and maximum depth.

\graphicspath{{figures/average_necessity}}

\graphicspath{{figures/supplimentary/}}

\section{Application to subreddit categories}\label{sec:subreddit}
As illustrated in Table \ref{tab:subreddit_group}, we have classified all subreddits into seven overarching categories, namely: `Funny', `Knowledge sharing', `Regional', `Sports', `Gaming', `Entertainment' and `Politics'. Notably, the `Gaming' and `Regional' categories include sub-categories for a more refined classification. Specifically, the `Gaming' category is divided into `General' and `Specific Games' sub-categories. The latter encompasses subreddits dedicated exclusively to individual games, such as `leagueoflegends', where discussions focus solely on that particular game. Similarly, the `Regional' category is subdivided into `General' and `With Restriction' sub-categories, with the latter including subreddits `AskUK,' `CasualUK,' and `KGBTR'. These three subreddits impose specific posting restrictions, requiring posts to be in the form of questions, non-political, and in the Turkish language, respectively. Fig.~\ref{fig:group_subreddit} illustrates the level of deliberation among subreddits in different categories, as measured by the `average width-average depth' (i.e., $P_{10}^a$), `maximum width-maximum depth' (i.e., $P_{10}^m$) and , `dyadic conversation count-average dyadic conversation length' metrics (i.e., $P_{10}^d$).

\subsection{The level of deliberation using `average width-average depth' metric}
Regarding the level of deliberation using `average width-average depth' metric (i.e., $P_{10}^a$), the subreddits within the `Sports' category generally exhibit a high degree of deliberation with the exception of the `golf' subreddit. Upon examination of the data related to posts within `golf' subreddit, we find a significant proportion of posts (26\%) are situated in Quadrant \uppercase\expandafter{\romannumeral4}, characterised by high average width but low average depth, compared to 12\% of posts in Quadrant II. We posit that this observation may be attributed to the `golf' subreddit's focus on users sharing personal experiences, which tends to be more casual and thus conducive to representation rather than in-depth argumentation. In contrast, other `Sports' subreddits predominantly center around discussions on professional games or players, fostering both high-level of argumentation and representation. Notably, there exists a subreddit named `ProGolf' (which did not fall within the top 90 hottest subreddits, so we did not collect data for that subreddit), where users primarily engage in discussions about professional golf games and players.

Within the `Regional' category, subreddits demonstrate varying levels of deliberation in the two sub-categories. Specifically, within the `General' sub-category, most subreddits exhibit exceedingly high levels of deliberation, ranking the highest among different categories, while the `With Restriction' sub-category tends to exhibit a lower degree of deliberation. Moreover, compared to the subreddit `unitedkingdom', a significant proportion of posts in the subreddits `AskUK' and `CasualUK' are situated in Quadrant \uppercase\expandafter{\romannumeral2}, indicating that many posts have high average width but lack high average depth. We conjecture this pattern is due to the nature of interactions within these subreddits. For example, in `AskUK' subreddit, a preliminary review of sample posts suggests that most comments directly respond to the original post's question, resulting in less substantive discussion that can be contributed to high depth. Fig. \ref{sfig:node_depth_distribuion_AskUK} and Fig.\ref{sfig:node_depth_distribuion_unitedkingdom} demonstrate the distribution of the depth of all comments within the subreddit `AskUK' and `unitedkingdom', respectively. Observing these figures, we find that a much larger proportion of comments within the `AskUK' subreddit have a depth of 1 (direct replies to original posts) compared to those within the `unitedkingdom' subreddit.


Turning to the `Gaming' category, subreddits also showcase diverse deliberation levels within the two sub-categories. Specifically, subreddits within the `General' sub-category display a notably higher degree of deliberation, while those within the `Specific Games' sub-category fall below the average of all categories. Upon examination of the data related to posts within each subcategory, we find that within `Specific Games' sub-category, a significant number of posts are situated in Quadrant \uppercase\expandafter{\romannumeral2} indicating that these posts possess high average depth but low average width. This observed difference may be attributed to the expansive range of ideas introduced in the discussion within `General' sub-category, fostering a greater representation of opinions. In contrast, the `Specific Games' sub-category focuses predominantly on discussions related to individual games, leading to more specialised and potentially narrower discourse. Consequently, this specialisation may result in a comparatively lower overall level of representation of opinions and thus lower level of deliberation.

The level of deliberation across all subreddits in the `Politics' category exceeds the average over all categories but does not reach the highest tier. After examining the data related to this category, we observe that a high proportion of posts are situated in Quadrant \uppercase\expandafter{\romannumeral2}, indicating a prevalent trend where posts exhibit high average depth but low width (see `worldnews' subreddit in Fig.~\ref{sfig:average_politics_plot}). For comparison, the same figure includes the subreddit `CFB' from the `Sport' category, which demonstrates a different trend and ranks highly in terms of deliberative potential.). In other words, the limiting factor preventing the `Politics' category from achieving top-tier deliberation levels is its average width. We conjecture that this phenomenon is driven by the nature of political discussions, which often require specialised knowledge and great effort to make a comment, resulting in less diverse representation.

Across the `knowledge sharing' and `Entertainment' categories, subreddits exhibit a diverse level of deliberation ranging from significantly above average to some of the lowest values. We attribute this diversity to the broad array of topics and nature within these subreddits. Although the overarching theme revolves around sharing knowledge and experiences, or entertainment, the subject matters explored are inherently varied, contributing to the observed range in deliberative engagement. For instance, a preliminary review of sample posts suggests that discussions within the `namenerds' subreddit tends to be more casual and relaxed compared to those within the `wallstreetbets' subreddit, resulting in the former experiencing less argumentation and thus lower level of deliberation.

\subsection{The level of deliberation using `maximum width-maximum depth' metric} \label{section:group_maximum}

\begin{figure} 
	\centering
	\includegraphics[width=0.9\linewidth]{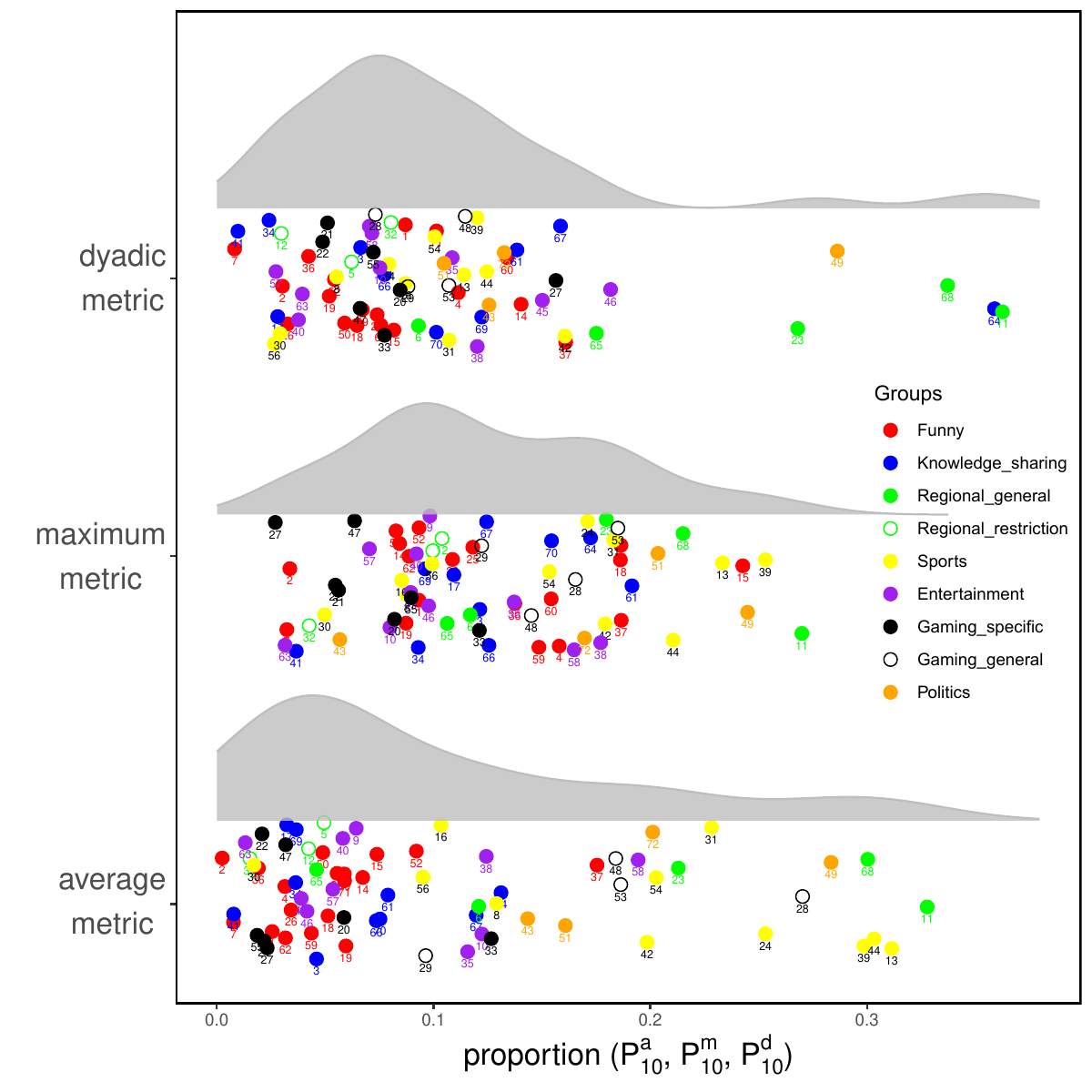 }
	\caption{Jitter plot with overlaid distribution curves to represent the degree of deliberation across various categories. The noise is added along the y-axis. The index numbers corresponding to each subreddit can be referenced in Table \ref{tab:subreddit_group}.
		 \vspace{-11pt}} 
	\label{fig:group_subreddit}
\end{figure}

Upon examining the `maximum width-maximum depth' metric (i.e., $P_{10}^m$), distinct patterns emerge among various subreddits when compared to the `average width-average depth' metric (i.e., $P_{10}^a$). This contrast is particularly pronounced within the `Funny' category. Specifically, as shown in Fig.~\ref{fig:group_subreddit}, subreddits within the `Funny' category demonstrate a significantly higher value in $P_{10}^m$ compared to $P_{10}^a$. The majority of subreddits in the `Funny' category demonstrate a noticeable lack of deliberation with regard to the `average width-average depth' metric (i.e., $P_{10}^a$), whereas many in this category surpass or closely approach the average level of deliberation when considering the `maximum width-maximum depth' metric (i.e., $P_{10}^m$), with some even achieving the highest levels.

\begin{figure*}	
 \subfigure []{
		\includegraphics[width=0.48\linewidth]{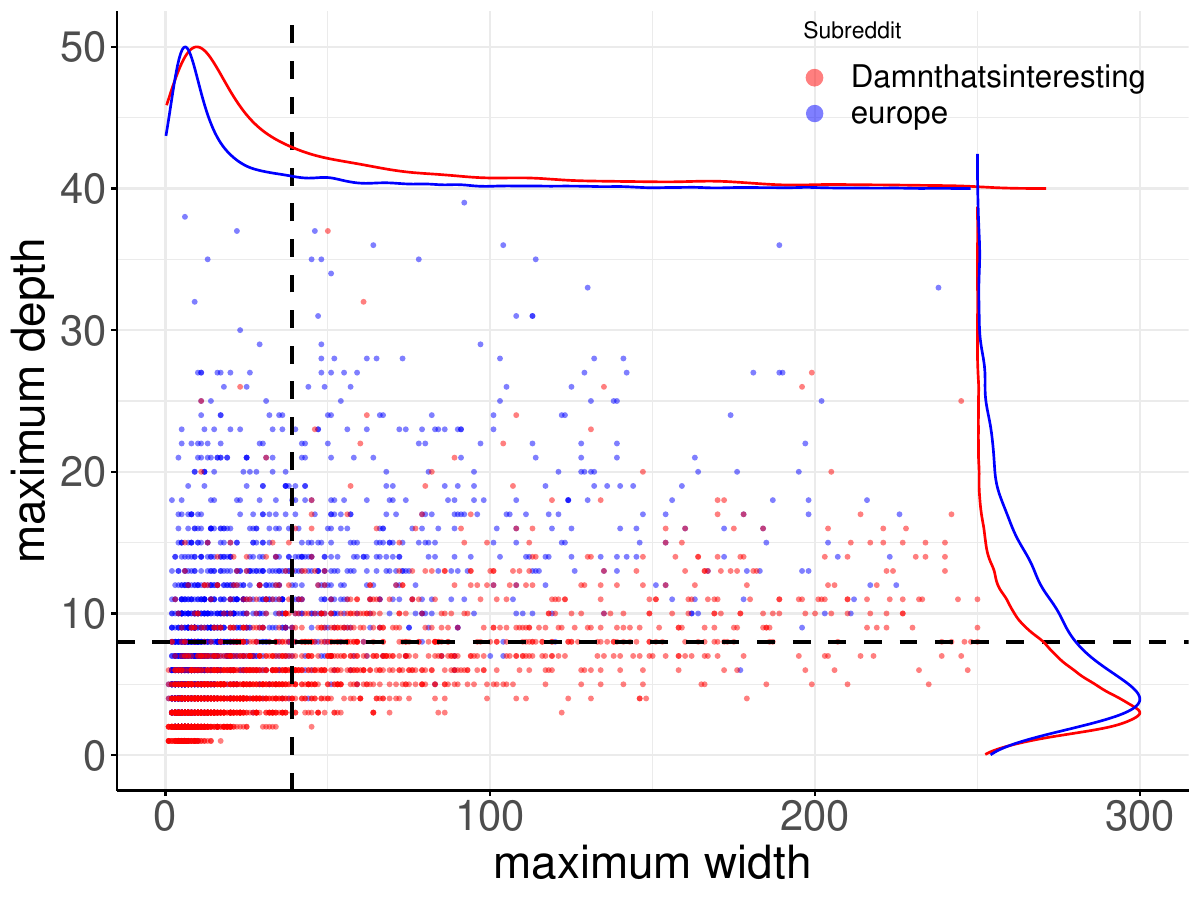}
		\label{sfig:maximum_two_scatter_plot}
  }\subfigure []{
		\includegraphics[width=0.48\linewidth]{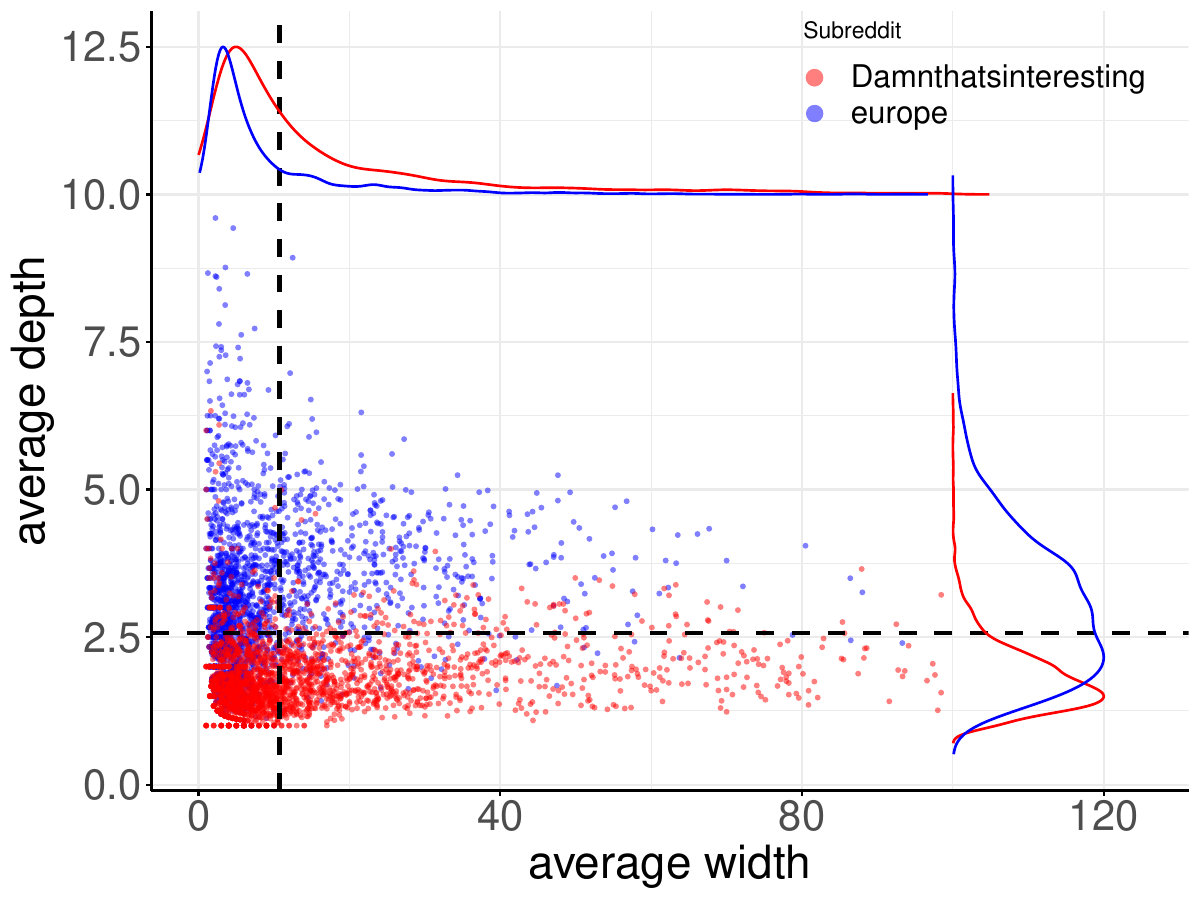}
		\label{sfig:average_two_scatter_plot}}
  \subfigure []{
		\includegraphics[width=0.48\linewidth]{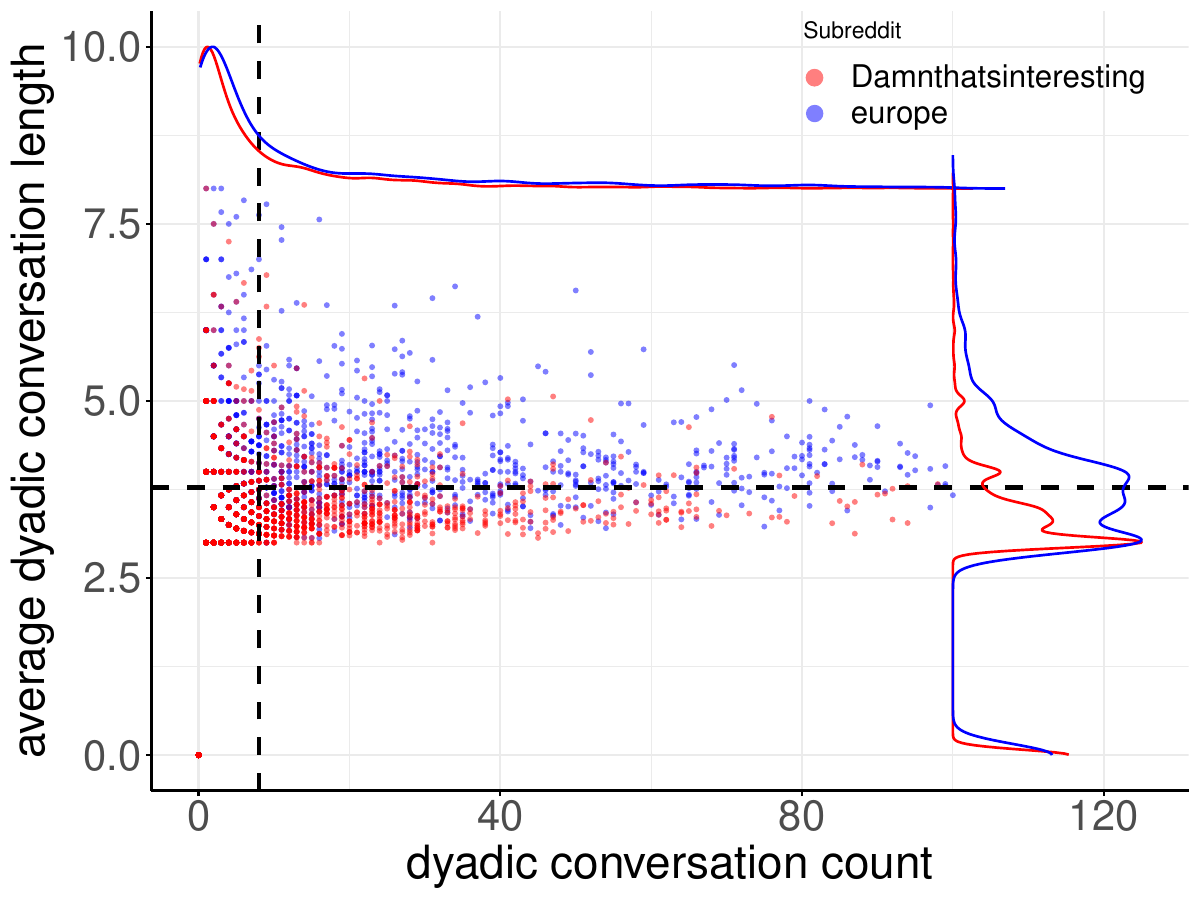}
		\label{sfig:dyadic_two_scatter_plot}
  }\subfigure []{
		\includegraphics[width=0.48\linewidth]{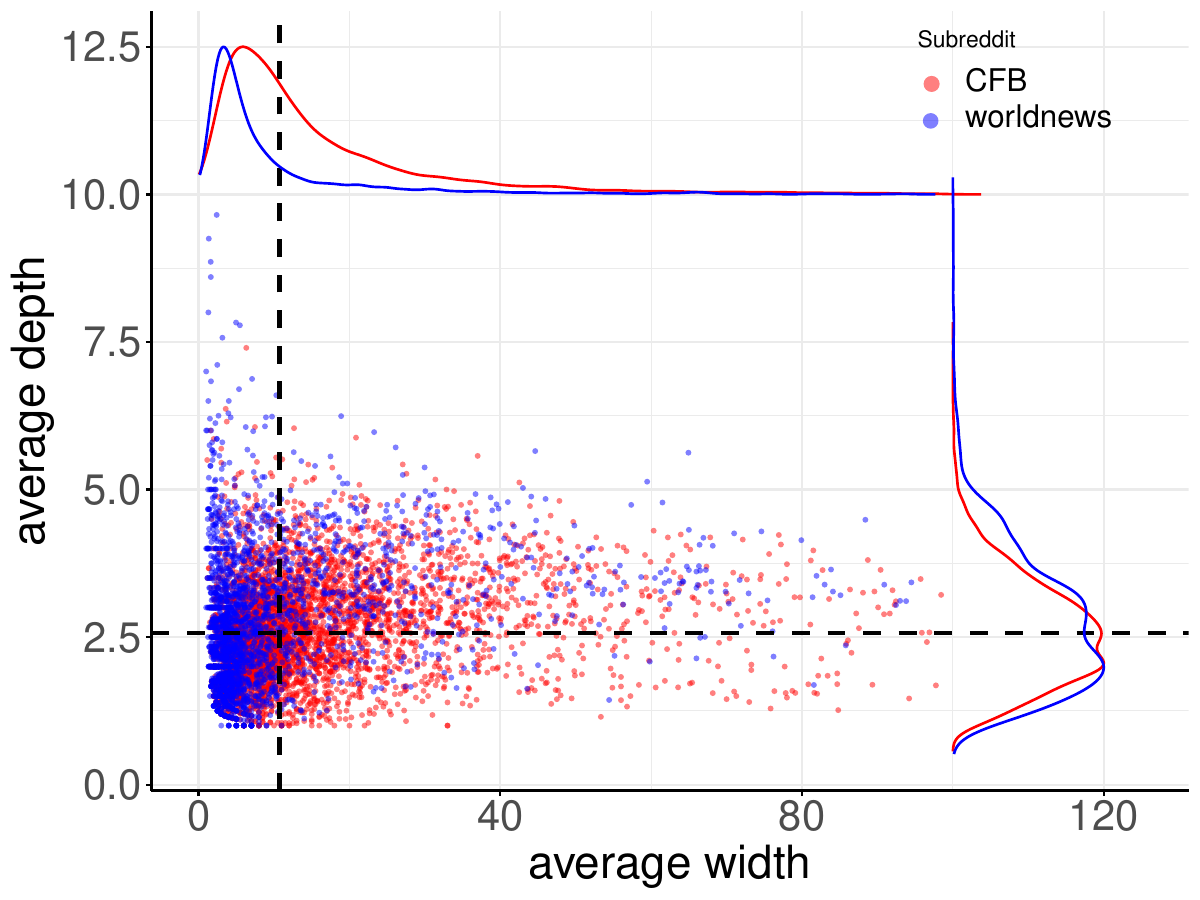}
		\label{sfig:average_politics_plot}}
  
  \caption{In panels (a), (b), and (c), we present visualisations of posts from the subreddits `Damnthatsinteresting' and `europe' using different coordinate systems: `maximum width-maximum depth', `average width-average depth', and `dyadic conversation count-average dyadic conversation length', respectively. The boundary lines defining the four quadrants are emphasised with dashed red lines. Panel (d) depicts posts from the subreddits `CFB' and `worldnews' plotted in the `average width-average depth' coordinate system.}
  
  \label{fig:funny_regional_scatter}
\end{figure*}

Fig.~\ref{fig:funny_regional_scatter} illustrates a scatter plot comparing posts from two representative subreddits, namely `Damnthatsinteresting' and `europe', within the `Funny' and `Regional' categories, across three different metrics (i.e., $P_{10}^a$, $P_{10}^m$ and $P_{10}^d$). Observations indicate that in respect to the `maximum width-maximum depth' metric (i.e., $P_{10}^m$), the proportion of posts situated in Quadrant \uppercase\expandafter{\romannumeral1} within these two subreddits is approximately the same. However, concerning the `average width-average depth' metric (i.e., $P_{10}^a$), the proportion of posts within the `Damnthatsinteresting' subreddit situated in Quadrant \uppercase\expandafter{\romannumeral1} is significantly lower than that of `europe' subreddit. Moreover, a large number of posts within `Damnthatsinteresting' subreddit occupy Quadrant \uppercase\expandafter{\romannumeral4} of the `average width-average depth' coordinate, indicating a prevailing trend where posts exhibit substantial average width but lack depth.

\begin{figure*}	
 \subfigure [AskUK]{
		\includegraphics[width=0.24\linewidth]{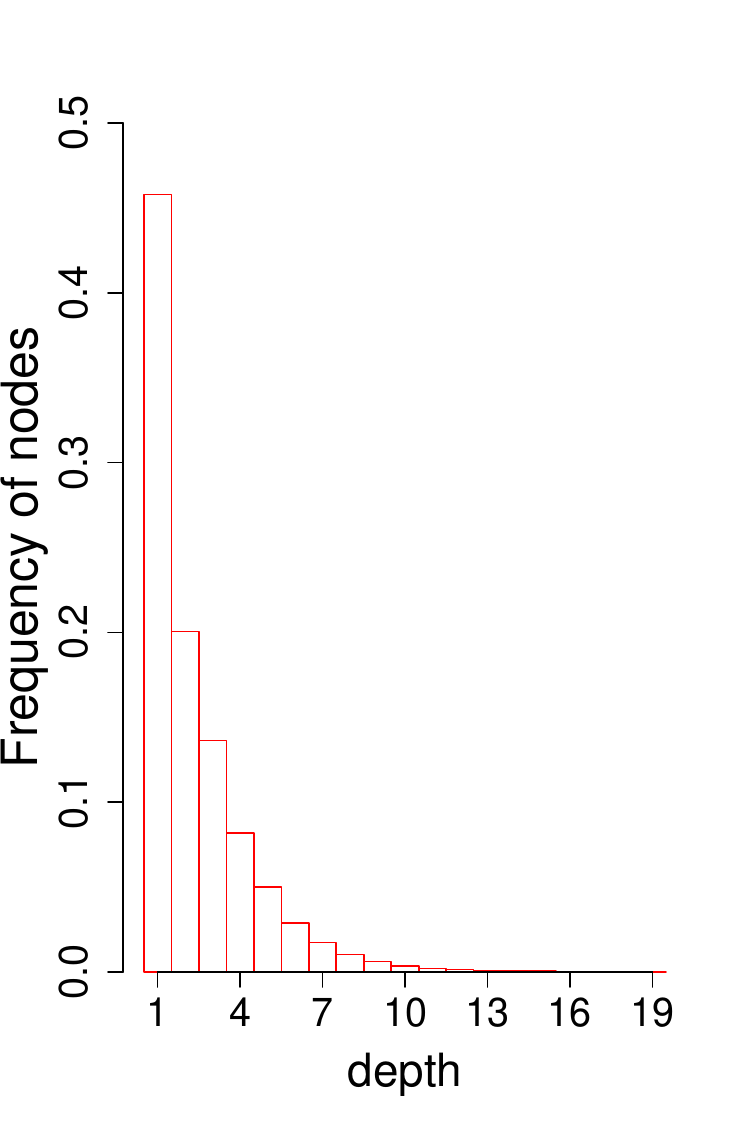}
		\label{sfig:node_depth_distribuion_AskUK}
  }\subfigure [unitedkingdom]{
		\includegraphics[width=0.24\linewidth]{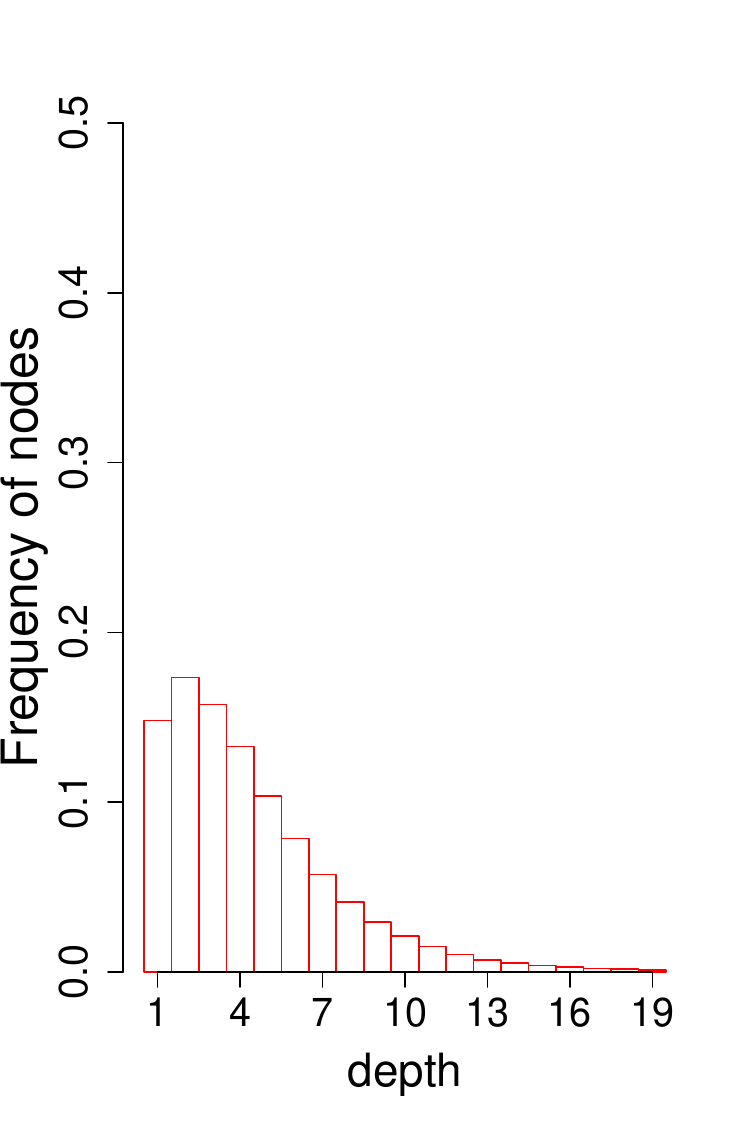}
		\label{sfig:node_depth_distribuion_unitedkingdom}}
 \subfigure [Damnthatsinteresting]{
		\includegraphics[width=0.24\linewidth]{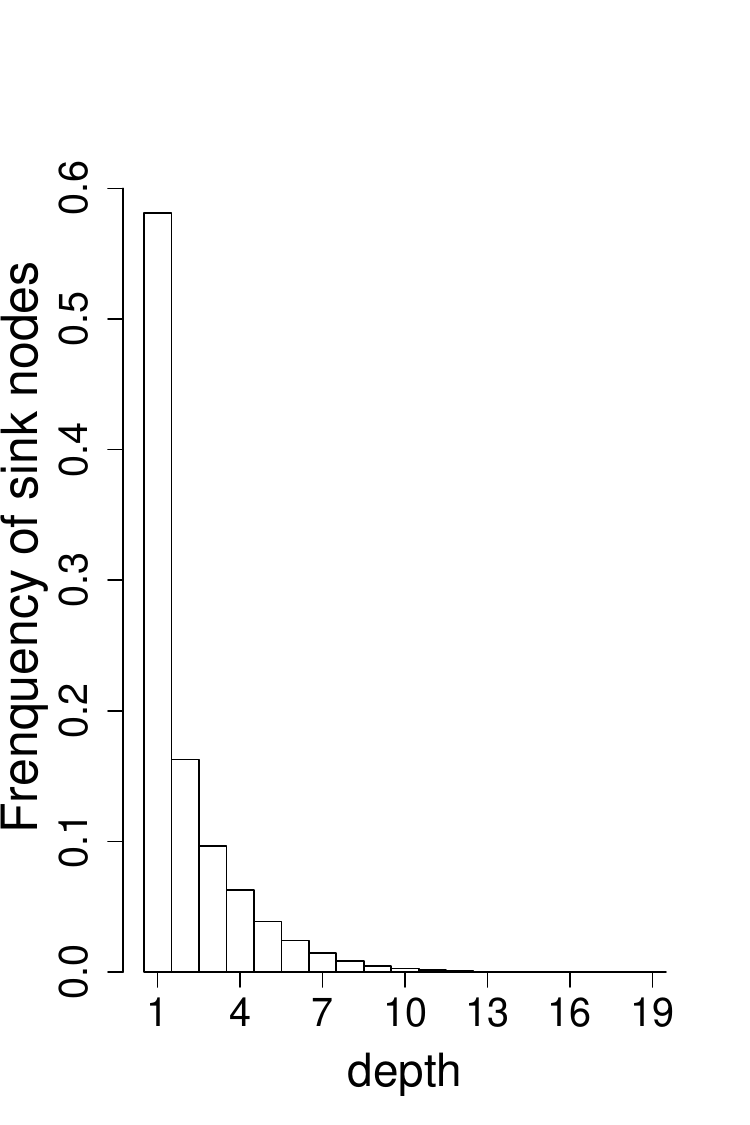}
		\label{sfig:sink_node_depth_distribuion_Damnthatsinteresting}
  }\subfigure [europe]{
		\includegraphics[width=0.24\linewidth]{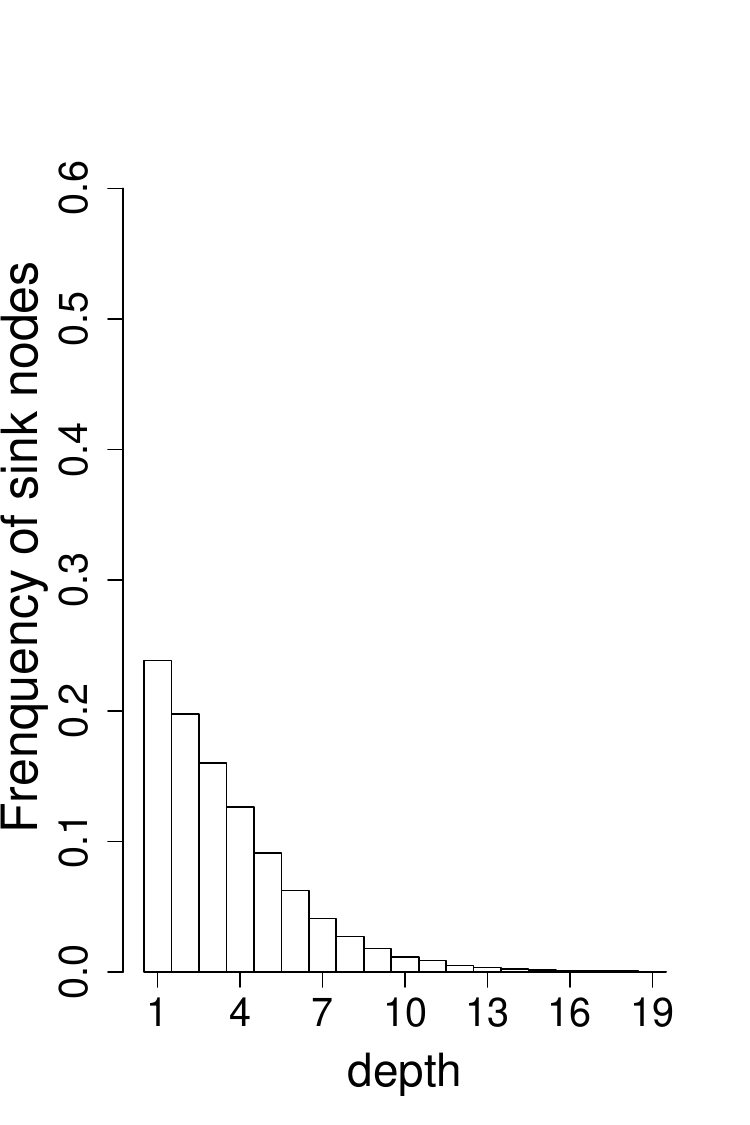}
		\label{sfig:sink_node_depth_distribuion_europe}}

  \caption{Comparison of the distribution of node depth and sink nodes' depth in different Subreddits. Panels (a) and (b) illustrate the distribution of all nodes within the subreddit `AskUK' and `unitedkingdom', respectively. Panels (c) and (d) illustrate the distribution of sink nodes within the subreddit `Damnthatsinteresting' and `europe', respectively.}
  \label{fig:depth_distribution}
\end{figure*}

Fig. \ref{sfig:sink_node_depth_distribuion_Damnthatsinteresting} and \ref{sfig:sink_node_depth_distribuion_europe} illustrates the distribution of sink nodes at various depths in the subreddits `Damnthatsinteresting' and `europe'. We observe that compared to `europe', the sink nodes within `Damnthatsinteresting' are predominantly concentrated at depth 1, with a sharp decline thereafter. This skewed depth distributions contributes to the disparity observed, where posts in the `Damnthatsinteresting' subreddit exhibit high maximum depth but low average depth. We attribute these observation to the inherent nature of the subreddits within the `Funny' category, where users tend to engage in a more casual manner, often resorting to passive consumption or providing direct comments on the original comment, rather than engaging in extended discussions.

The divergence in deliberation levels between the `maximum width-maximum depth' (i.e., $P_{10}^m$) and `average width-average depth' metrics (i.e., $P_{10}^a$) highlights the utility of the latter metric in offering additional insights into deliberation, particularly when the discussion tree shows significant skew.

\subsection{The level of deliberation using `dyadic conversation count-average dyadic conversation length' metric}

Within `Sports', `Regional', `knowledge sharing', `Entertainment' and `Politics' category, we observe similar trend in the level of deliberation regarding `average width-average depth' (i.e., $P_{10}^a$) and `dyadic conversation count-average dyadic conversation length' metric (i.e., $P_{10}^d$). However, there are also notable differences between these two metrics in the `Gaming' and `Funny' category.

Within `Gaming' category, regarding the dyadic conversation metric (i.e., $P_{10}^d$), there is no significant difference observed in the proportion of posts situated in Quadrant \uppercase\expandafter{\romannumeral1} between the two sub-categories. This suggests that the degree of deliberation within these two sub-categories is similar. In this regard, most subreddits within `Gaming' category exhibit a degree of deliberation around the average level. Nevertheless, compared to `General Games' sub-category,  the `Specific Games' sub-category contains a higher proportion of posts situated in Quadrant \uppercase\expandafter{\romannumeral2}. This signifies that many posts have high average dyadic conversation length but low in dyadic conversation count. In other words, the limited dyadic conversation count prevents the `Specific Games' sub-category from achieving a higher degree of deliberation, likely due to the specialised and narrow focus of discussions within these subreddits. However, once reciprocal discussions begin, the specialised nature of the discussions can lead to relatively lengthy interactions.

On the contrary, within the `General Games' sub-category, a large proportion of posts are situated in Quadrant \uppercase\expandafter{\romannumeral4}, indicating that many posts have high dyadic conversation count but lack sufficient average dyadic conversation length. This observation may be attributed to the broader discussion topics in this sub-category, which tend to encourage diverse viewpoints and, consequently, high dyadic conversation counts. However, the length of these conversations may be limited due to the less in-depth discussions.

Within the `Funny' category, certain subreddits exhibit a high level of deliberation regarding `dyadic conversation count-average dyadic conversation length' metric (i.e., $P_{10}^d$). As discussed in Section~\ref{section:group_maximum}, it was explained that all subreddits within the `Funny' category demonstrate a low level of deliberation according to the `average width-average depth' metric ($P_{10}^a$) due to low average depth. This implies that while most posts within the `Funny' category lack substantive argumentation (average depth), there is still a strong likelihood of engaging in decent-length reciprocal discussions. This observation suggests that a high level of deliberation from the prospective of dyadic conversation metrics does not necessarily need high average depth.

As another example, within subreddit `OnePiecePowerScaling', the level of deliberation regarding `dyadic conversation count-average dyadic conversation length' metric (i.e., $P_{10}^d$) ranks at the top tier, with $P_{10}^d=0.18$. In contrast, regarding `average width-average depth' metric (i.e., $P_{10}^a$), the degree of deliberation within that subreddit ranks among the lowest tiers, with $P_{10}^a=0.044$, primarily due to its modest average width. This observation suggests that a high level of deliberation from the prospective of dyadic conversation metrics does not necessarily need high average width, either.

\section{Actor layer network analysis}\label{sec:actor_layer}

As outlined in Section~\ref{Sec:two_layer_model}, we constructed a distinct actor layer network for each subreddit based on user-reply interactions across all collected posts on that subreddit. When researching deliberation within a subreddit, our primary interest lies in studying the behaviour of the highly active users who engaged in that subreddit regularly. Thus, we refine the actor-layer networks of subreddits by excluding inactive users who contribute fewer than 3 comments in the corresponding subreddits within the 8-week collection period. Additionally, this refinement can also reduce the computation burden. On average, 64.3\% of nodes and 26.8\% of edges were removed from the actor-layer network for a given subreddit during this refinement process. We then compute the structural properties of these refined actor layer networks, incorporating both the direction and weight of edges. Table~\ref{table:actor_layer_average_property} shows the mean and 95\% confidence interval of the computed properties.

\begin{table}[!ht]
    \centering

    \begin{tabular}{l l l l}
    \hline
        Properties & Description & Mean value & Confidence interval \\ \hline
        $n_r$ &Number of nodes& 15,417 & [13,728, 17,107] \\ 
        $m_r$ &Number of edges& 153,270 & [131,195, 175,344] \\ 
        $F_g$ &Percentage of nodes in the giant component& 78.0\% & [75.0\%, 80.9\%] \\ 
        $l$ & Average shortest path length & 4.55 & [4.33, 4.78] \\ 
        $C$ &Global clustering coefficient &0.0583 & [0.0478, 0.0688] \\ 
        $d$ &Degree assortativity &-0.0126 & [-0.0352, 0.01] \\ 
        $\alpha_{in}$ & Power law coefficient for in-degree& 2.399 & [2.296, 2.503] \\ 
        $\alpha_{out}$ &Power law coefficient for out-degree& 2.903 & [2.812, 2.995] \\     
        \hline
    \end{tabular}
    \caption{Mean value and 95\% confidence interval of the structural properties of the refined actor layer networks}
    \label{table:actor_layer_average_property}
\end{table}

\subsection{General properties}

\subsubsection{Sparse topology and giant component}
As illustrated in Table~\ref{table:actor_layer_average_property}, the refined actor layer networks across 72 subreddits have an average of $n_r=15,417$ nodes and $m_r=153,270$ edges, with 95\% confidence interval of [13,728, 17,107] and [131,195, 175,344], respectively. Notably, the average edge count $m_r$ is significantly lower than the total number of potential edges in a network of size $n_r$ (i.e., $\mathcal{O}(n_r^2)$), indicating a pronounced sparsity within the refined networks. In addition, for each refined actor layer network, we identify a giant component by extracting the largest strongly connected sub-graph of the original graph. On average, across the 72 subreddits, the giant component contains $F_g=78.0\%$ of the actors from the refined network, with a 95\% confidence interval of [75.0\%, 80.9\%]. This suggests that for most subreddit, a substantial majority of actors coalesce into a giant component, leaving only a minor fraction of actors disconnected from it. A preliminary review of network data suggests that these disconnected actors typically form pairs or, at most, small clusters of up to five actors. In the following, all structural properties are computed based on the giant component of the refined networks for the 72 subreddits.

\subsubsection{Small world characteristics}
We adopted the methodology proposed by Watts and Strogatz \citep{watts1998smallworld} to determine whether actor layer networks of 72 subreddits exhibit `small world' characteristics. Based on their method, a network is qualified as `small world' if it meets the following two criteria. Firstly, its global clustering coefficient must be significantly higher than that of a random network with the same number of nodes, sparsity and edge weights. Secondly, its mean shortest distance between nodes should be similar to that of the corresponding random network.

We calculate the average shortest distance between nodes and the global clustering coefficient for the giant component of each refined network. We then used the Erdos-Renyi model to construct 100 directed weighted random networks, matching the number of nodes, edges, and edge weights to the mean values of these parameters from the giant component of each of the 72 refined networks. The average shortest distance between nodes and the global clustering coefficient are also computed for these 100 random networks. For the 72 giant components, as illustrated in Table~\ref{table:actor_layer_average_property}, the mean value of the average shortest distance between nodes is $d=4.55$ (with a 95\% confidence interval of [4.33, 4.78]), which is similar to that of the 100 random networks (mean value of 4.39). Meanwhile, the mean value of the global clustering coefficient for the 72  giant components is $C=0.0583$ (with a 95\% confidence interval of [0.0478, 0.0688]), which is approximately two orders of magnitude greater than that of the 100 random networks (mean value of 0.00092). This indicates the giant components of the refined networks possess small world characteristics.

\subsubsection{Degree assortativity}
Another important property of social networks is the degree assortativity coefficient, which measures the tendency for nodes with similar degree to be connected. We compute this coefficient for the giant component of the 72 refined networks following the methodology outlined by \cite{yuan2021assortativity}, which also accounts for both edge direction and weight. As illustrated in Table~\ref{table:actor_layer_average_property}, the computed mean coefficient value, along with its 95\% confidence interval, is $d=-0.0126$ and [-0.0352, 0.010], respectively, which is far from $\pm 1$. This finding suggests that the majority of the giant component of the 72 refined networks neither exhibit pronounced assortative nor disassortative behaviour. Essentially, users do not demonstrate a strong inclination to engage with comments from users of similar or different degrees. This observed behavioural pattern contrasts with that of real-world social networks and user-based platforms like Twitter or Facebook, where the reply-interaction networks typically demonstrate strong disassortativity \citep{wang2017detecting}. Similar trends have been identified in related studies, emphasising inherent disparities between real-world networks and large-scale content-based online social networks \citep{gomez2008statistical,HOLME2004155}.

We hypothesise that the observed difference in assortative behaviour of actor layer networks between Reddit (which is content-based platform) and user-based platforms is influenced by the distinct nature of user exposure in these platforms. In Reddit, users subscribe to specific subreddits, and their exposure to content is shaped by the collective activity and rules within those subreddits. This subscription model can lead users to encounter comments and posts from a diverse range of users, including both active and less active users. Conversely, in user-based platforms, such as Twitter or Facebook, users typically interact more closely with comment shared by the users they actively follow. This tends to create a more selective exposure to comment, with a higher likelihood of engaging primarily with comment from more active users who typically have larger followings. The exposure pattern on Reddit, characterised by broader and less selective exposure to various users' contributions within subscribed subreddits, may contribute to the observed lack of assortative behaviours in actor layer networks. This contrast highlights how platform-specific features, such as subscription models and follower models, can significantly impact network dynamics within online social networks. 

\subsubsection{Degree distribution}
The final property we analyse for the giant component of the refined actor layer network is the degree distribution of nodes. We observed that the degree distributions (both in- and out-degree) for all 72 subreddits exhibit a power-law pattern, suggesting a significant level of heterogeneity of degree within these subreddits. We then use the method developed by \cite{power_law_fit_Clauset}, implemented in the `poweRlaw’ R package to compute the corresponding power law coefficient for these 72 giant components. As illustrated in Table~\ref{table:actor_layer_average_property}, the mean values of these power law coefficient for in-degree and out-degree, respectively, are $\alpha_{in}=2.399$ and $\alpha_{out}=2.903$, with 95\% confidence interval of [2.296, 2.503] and [2.812, 2.995]. These coefficients indicate the rate at which the frequency of nodes decreases as their degree increases. Specifically, an in-degree coefficient of $\alpha_{in}=2.399$  suggests that the network has a few nodes with very high in-degrees, while the majority have low in-degrees, with a similar finding for out-degree ($\alpha_{out}=2.903$). The relatively large in-degree and out-degree power law coefficients reveal the presence of a small number of highly influential users that receive a lot of replies or make a lot of comments. The difference between the power law coefficients for in-degree and out-degree indicate that the former distribution has a heavier tail than the latter, which is consistent with findings for reply distributions on Twitter \citep[see, e.g.,][] {Federico_etal2024_multiscale}. This suggests that even though it is less common for people to follow one another on Reddit, the comment upvote and karma systems still lead to cumulative advantage, with a few users garnering large amounts of attention. In contrast, the fact that writing comments requires effort and time means that we observe relatively fewer ``superstar commenters'', based on the number of comments authored.  


\subsection{Highly active users}
\begin{figure} 
	\centering
	\includegraphics[width=0.9\linewidth]{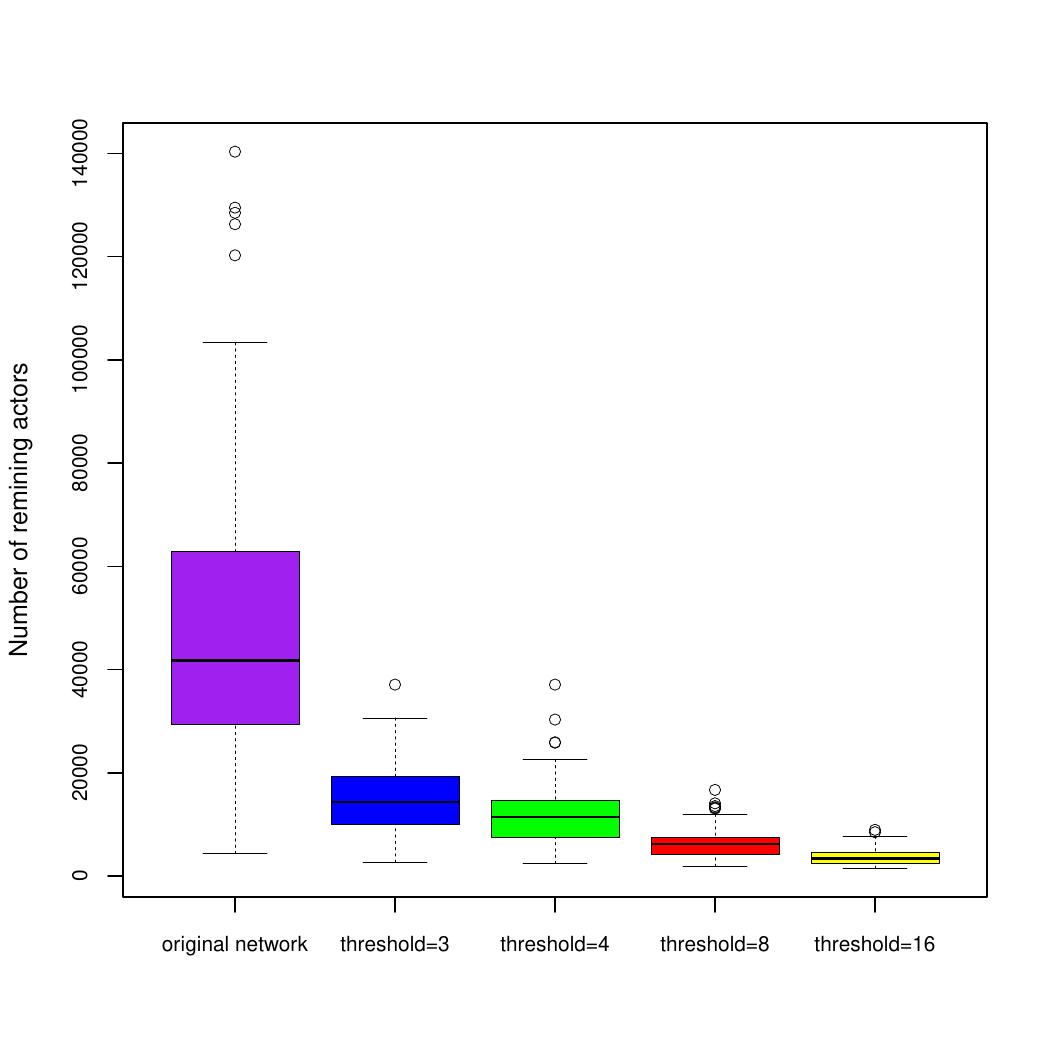}
	\caption{Box and whisker plot illustrating the count of remaining users based on distinct refinement threshold. The threshold, progressing from left to right in the figure, involve the exclusion of occasional users contributing fewer than 0, 3, 4, 8, and 16 comments within the 8-week period. \vspace{-11pt} }
	\label{fig:boxplot}
\end{figure}

Further refinement for the actor layer networks of subreddits is carried out by removing occasional users with less than 4, 8, and 16 comments within the 8-weeks period from the actor layer network. A box and whisker plot illustrating the number of remaining users after each refinement is presented in Fig.~\ref{fig:boxplot}. We observe that for refined networks where every user contributes more than or equal to 16 comments, the median network size is 3798, with a quartile range of [2,479, 4,483]. 

This refinement provides an interesting finding as follows. First, note that there is significant variation across subreddits in the number of subscribers. For example, the subreddit `MadeMeSmile' has 9.6 million subscribers which is about 60 times more than that of the subreddit `neoliberal', with 166,000 subscribers. The number of actors in the original network (left box-and-whisker) also varies substantially (with a range of over 120,000 when taking into account outliers). As we refine the network only retaining users who contribute more than a certain number of comments, the range in the network size shrinks rapidly. 
Both r/MadeMeSmile and r/neoliberal have around 2,000 users that contributed 16 comments or more. Our findings suggest that, regardless of the number of subscribers and casual users, the number of highly active users across all 72 subreddits is relatively similar. Note that this does not imply that the highly active users of one subreddit are the same as those of another subreddit, but rather, suggests an upper limit to how many users contribute regularly to the discussions in a subreddit. Future research will involve investigating overlaps in the highly active users across subreddits.


\graphicspath{{figures/box_whisker/}}

\subsection{Relationship between the characteristics of actor layer and deliberation}
\begin{figure} 
	\centering
	\includegraphics[width=0.9\linewidth]{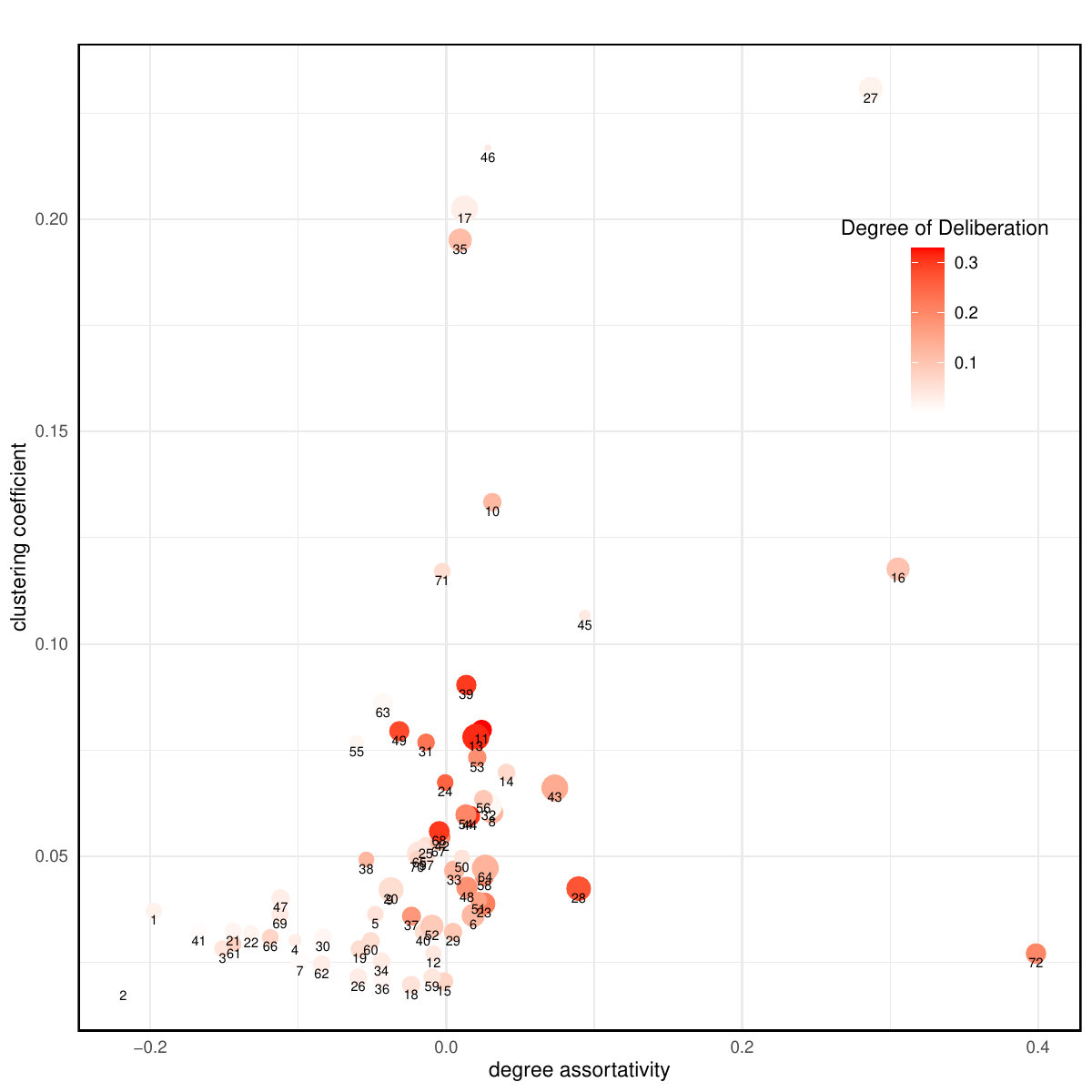}
	\caption{Heat map for the correlation of degree assortativity, power law coefficient, global clustering coefficient, and the level of deliberation measured by `average width-average depth' metric (i.e., $P_{10}^a$). The x-axis and y-axis represent degree assortativity and global clustering coefficient, respectively. The size of the point is inversely proportional to the power law coefficient and the color of the points captures the degree of deliberation.   \vspace{-11pt} }
	\label{fig:hotmap}
\end{figure}

We investigate the correlation between properties of the giant components of the refined networks whose users contribute more than or equal to 3 comments and the level of deliberation within subreddits. Through extensive observation, we identify that the degree assortativity, the power-law coefficient, and the global clustering coefficient tend to correlated with the level of deliberation of subreddits. In Fig. \ref{fig:hotmap}, we present a heat-map illustrating the relationships among these three attributes and the level of deliberation measured by the `average width-average depth' metric (i.e., $P_{10}^a$) for all 72 subreddits. Each point on the scatter plot represents a subreddit, with the size of the point inversely proportional to the value of the in-degree power-law coefficient. The color of each point indicates the level of deliberation (i.e., $P_{10}^a$), with deeper shades of red signifying higher degrees of deliberation. 

It is evident that the upper-right area, characterised by larger point sizes, exhibits deeper colors, indicating a higher level of deliberation. This observation suggests that in subreddits where the in-degree distribution is more homogeneous, actors tend to engage with others of similar degree (positive degree assortativity), and such actors are closely clustered, are associated with higher levels of deliberation. We hypothesise that the positive correlation between the degree assortativity and the level of deliberation arises from a prevailing trend in online discussion forums. First, we remark that an increase in degree assortativity is most likely to arise when nodes with low in-degree respond to other nodes with low in-degree, as such nodes are far more numerous than nodes with high in-degree. Second, we hypothesise that high in-degree nodes (those that receive many responses) do not reply as frequently as they receive responses, and are selective as to who they reply to, as observed in other social media platforms~\citep{xiong2020understanding}. For example, within the subreddit `MadeMeSmile', users with an in-degree greater than 194 attract 50\% of replies but contribute only 1.5\% of comments. Putting these two points together, we conjecture that the level of deliberation increases when low in-degree nodes respond to other low in-degree nodes (and thus facilitating broader argumentation and representation) rather than responding to high in-degree nodes (who are unlikely to reply and thus hampers the discussion).


Similarly, we have generated a heat-map illustrating the relationships among the attributes of the giant components of the refined networks and the level of deliberation measured by the `maximum width-maximum depth' metric (i.e., $P_{10}^m$) and `dyadic conversation count-average dyadic conversation length' metric (i.e., $P_{10}^d$), respectively. However, in our analysis, we have observed that these factors exhibit a weaker correlation.

\section{Conclusions}\label{sec:conclusions}

In this paper, we presented a two-layer network model that can be used to capture discussions that unfold on online platforms such as Reddit, and proposed associated network metrics to capture the level of deliberation in a discussion. We applied this network model and the associated metrics to a large Reddit dataset spanning 72 different subreddits, reporting several interesting findings on differences and similarities between subreddits across different categories. 

While this paper has presented a model and demonstrated its application to Reddit, there are a number of directions for future research. First, we should further examine the interlayer dependencies to identify additional insights into deliberation; currently, we have only used the interlayer links to define the dyadic conversation metrics. Second, our analysis focuses on the discussion thread at the end (i.e. when virtually all comments have been posted). It would be insightful to examine the temporal dynamics, including how the various metrics evolve over time. This might allow us to understand which discussion trajectories are most likely to end up in a Type I network structure. Third, the role of high-degree users, both in-degree and out-degree, in facilitating deliberation should be further investigated. Finally, application of our model and the associated metrics to other online discussion platforms could reveal how different platform affordances might lead to differences in the discussion structures, and consequently, its deliberative potential.


\bibliography{pp}

\begin{thebibliography}{}

\bibitem[Ackerman and Fishkin, 2002]{ackerman2002deliberation}
Ackerman, B. and Fishkin, J. (2002).
\newblock Deliberation day.

\bibitem[Ackland et~al., 2023]{ackland2023reciprocal}
Ackland, R., Gumbert, F., P{\"u}tz, O., Gertzel, B., and Orlikowski, M. (2023).
\newblock Reciprocal communication and political deliberation on twitter.
\newblock {\em Social Sciences}, 13(1):5.

\bibitem[Arag{\'o}n et~al., 2017]{Argon_2017_sentiment}
Arag{\'o}n, P., Kaltenbrunner, A., Calleja-L{\'o}pez, A., Pereira, A., Monterde, A., Barandiaran, X.~E., and G{\'o}mez, V. (2017).
\newblock Deliberative platform design: The case study of the online discussions in decidim barcelona.
\newblock In {\em Social Informatics: 9th International Conference, SocInfo 2017, Oxford, UK, September 13-15, 2017, Proceedings, Part II 9}, pages 277--287. Springer.

\bibitem[Clauset et~al., 2007]{power_law_fit_Clauset}
Clauset, A., Shalizi, C., and Newman, M. (2007).
\newblock Power-law distributions in empirical data.
\newblock {\em SIAM Review}, 51.

\bibitem[Dahiya et~al., 2021]{dahiya2021hates}
Dahiya, S., Sharma, S., Sahnan, D., Goel, V., Chouzenoux, E., Elvira, V., Majumdar, A., Bandhakavi, A., and Chakraborty, T. (2021).
\newblock Would your tweet invoke hate on the fly? forecasting hate intensity of reply threads on twitter.
\newblock In {\em Proceedings of the 27th ACM SIGKDD Conference on Knowledge Discovery \& Data Mining}, pages 2732--2742.

\bibitem[Dryzek, 1994]{dryzek1994australian}
Dryzek, J.~S. (1994).
\newblock Australian discourses of democracy.
\newblock {\em Australian Journal of Political Science}, 29(2):221--239.

\bibitem[Elster, 1998]{elster1998deliberative}
Elster, J. (1998).
\newblock {\em Deliberative democracy}, volume~1.
\newblock Cambridge University Press.

\bibitem[Federico et~al., 2024]{Federico_etal2024_multiscale}
Federico, L., Mounim, A., Caldarelli, G., and Riotta, G. (2024).
\newblock Multi-scale analysis of the community structure of the twitter discourse around the italian general elections of september 2022.
\newblock {\em Scientific Reports}, 14(1):15980.

\bibitem[Gertzel et~al., 2024]{vosonSML}
Gertzel, B., Ackland, R., Graham, T., and {Bórquez Vivanco}, M.~F. (2024).
\newblock {vosonSML}: Collecting social media data and generating networks for analysis.
\newblock \url{https://cran.r-project.org/package=vosonSML}.

\bibitem[G{\'o}mez et~al., 2008]{gomez2008statistical}
G{\'o}mez, V., Kaltenbrunner, A., and L{\'o}pez, V. (2008).
\newblock Statistical analysis of the social network and discussion threads in slashdot.
\newblock In {\em Proceedings of the 17th international conference on World Wide Web}, pages 645--654.

\bibitem[Gonzalez-Bailon et~al., 2010]{gonzalez2010structure}
Gonzalez-Bailon, S., Kaltenbrunner, A., and Banchs, R.~E. (2010).
\newblock The structure of political discussion networks: a model for the analysis of online deliberation.
\newblock {\em Journal of Information Technology}, 25(2):230--243.

\bibitem[Guerrero-Sol{\'e}, 2018]{guerrero2018twitter_actor}
Guerrero-Sol{\'e}, F. (2018).
\newblock Interactive behavior in political discussions on twitter: Politicians, media, and citizens’ patterns of interaction in the 2015 and 2016 electoral campaigns in spain.
\newblock {\em Social Media+ Society}, 4(4):2056305118808776.

\bibitem[Habermas, 1985]{habermas1985remarks}
Habermas, J. (1985).
\newblock Remarks on the concept of communicative action.
\newblock In {\em Social action}, pages 151--178. Springer.

\bibitem[Holme et~al., 2004]{HOLME2004155}
Holme, P., Edling, C.~R., and Liljeros, F. (2004).
\newblock Structure and time evolution of an internet dating community.
\newblock {\em Social Networks}, 26(2):155--174.

\bibitem[Howard and Parks, 2012]{howard2012socialmedia_review_paper}
Howard, P.~N. and Parks, M.~R. (2012).
\newblock Social media and political change: Capacity, constraint, and consequence.

\bibitem[Kizilhan and Kizilhan, 2016]{kizilhan2016rise_polarisation}
Kizilhan, T. and Kizilhan, S.~B. (2016).
\newblock The rise of the network society-the information age: Economy, society, and culture.
\newblock {\em Contemporary Educational Technology}, 7(3):277--280.

\bibitem[Medvedev et~al., 2019a]{Medvedev_overview_Reddit}
Medvedev, A., Lambiotte, R., and Delvenne, J.-C. (2019a).
\newblock {\em The Anatomy of Reddit: An Overview of Academic Research}, pages 183--204.

\bibitem[Medvedev et~al., 2019b]{medvedev2019modelling}
Medvedev, A.~N., Delvenne, J.-C., and Lambiotte, R. (2019b).
\newblock Modelling structure and predicting dynamics of discussion threads in online boards.
\newblock {\em Journal of Complex Networks}, 7(1):67--82.

\bibitem[Mendoza et~al., 2023]{mendoza2023study_misinformation}
Mendoza, M., Valenzuela, S., N{\'u}{\~n}ez-Mussa, E., Padilla, F., Providel, E., Campos, S., Bassi, R., Riquelme, A., Aldana, V., and L{\'o}pez, C. (2023).
\newblock A study on information disorders on social networks during the chilean social outbreak and covid-19 pandemic.
\newblock {\em Applied Sciences}, 13(9):5347.

\bibitem[Nishi et~al., 2016]{nishi2016twitterreply}
Nishi, R., Takaguchi, T., Oka, K., Maehara, T., Toyoda, M., Kawarabayashi, K.-i., and Masuda, N. (2016).
\newblock Reply trees in twitter: data analysis and branching process models.
\newblock {\em Social network analysis and mining}, 6:1--13.

\bibitem[Noveck, 2009]{noveck2009oneline_forum}
Noveck, B.~S. (2009).
\newblock {\em Wiki government: How technology can make government better, democracy stronger, and citizens more powerful}.
\newblock Brookings Institution Press.

\bibitem[Price, 2009]{price2009citizens}
Price, V. (2009).
\newblock Citizens deliberating online: Theory and some evidence.
\newblock {\em Online deliberation: Design, research, and practice}, pages 37--58.

\bibitem[Romsdahl, 2005]{romsdahl2005political}
Romsdahl, R.~J. (2005).
\newblock Political deliberation and e-participation in policy-making.
\newblock {\em CLCWeb: Comparative Literature and Culture}, 7(2):7.

\bibitem[Shane, 2004]{shane2004democracy}
Shane, P.~M. (2004).
\newblock {\em Democracy online: the prospects for political renewal through the Internet}.
\newblock Psychology Press.

\bibitem[Steibel and Estevez, 2015]{steibel2015designing}
Steibel, F. and Estevez, E. (2015).
\newblock Designing web 2.0 tools for online public consultation.
\newblock {\em Impact of information society research in the global south}, pages 243--263.

\bibitem[Ulu{\c{c}} et~al., 2010]{ulucc2010blogs_political}
Ulu{\c{c}}, G., Yilmaz, M., and Isikdag, U. (2010).
\newblock Blogs and forums in a presidential election process in turkey.
\newblock In {\em Handbook of research on social interaction technologies and collaboration software: Concepts and trends}, pages 372--382. IGI Global.

\bibitem[Wang et~al., 2017]{wang2017detecting}
Wang, T., Brede, M., Ianni, A., and Mentzakis, E. (2017).
\newblock Detecting and characterizing eating-disorder communities on social media.
\newblock In {\em Proceedings of the Tenth ACM International conference on web search and data mining}, pages 91--100.

\bibitem[Watts and Strogatz, 1998]{watts1998smallworld}
Watts, D.~J. and Strogatz, S.~H. (1998).
\newblock Collective dynamics of ‘small-world’networks.
\newblock {\em nature}, 393(6684):440--442.

\bibitem[Wilson et~al., 2012]{wilson2012review_facebook}
Wilson, R.~E., Gosling, S.~D., and Graham, L.~T. (2012).
\newblock A review of facebook research in the social sciences.
\newblock {\em Perspectives on psychological science}, 7(3):203--220.

\bibitem[Xiong et~al., 2020]{xiong2020understanding}
Xiong, J., Feng, X., and Tang, Z. (2020).
\newblock Understanding user-to-user interaction on government microblogs: An exponential random graph model with the homophily and emotional effect.
\newblock {\em Information Processing \& Management}, 57(4):102229.

\bibitem[Yuan et~al., 2021]{yuan2021assortativity}
Yuan, Y., Yan, J., and Zhang, P. (2021).
\newblock Assortativity measures for weighted and directed networks.
\newblock {\em Journal of Complex Networks}, 9(2):cnab017.

\bibitem[Zimmer and Proferes, 2014]{zimmer2014topology}
Zimmer, M. and Proferes, N.~J. (2014).
\newblock A topology of twitter research: disciplines, methods, and ethics.
\newblock {\em Aslib Journal of Information Management}, 66(3):250--261.

\end{thebibliography}
\bibliographystyle{apalike}

\end{document}